\documentstyle[preprint,prb,aps]{revtex}
\begin{document}
\draft
\def\gappeq{\mathrel{\rlap {\raise.5ex\hbox{$>$}}
{\lower.5ex\hbox{$\sim$}}}}

\def\lappeq{\mathrel{\rlap{\raise.5ex\hbox{$<$}}
{\lower.5ex\hbox{$\sim$}}}}

\def \gsim{\lower.8ex\hbox{$\sim$}\kern-.75em\raise.45ex\hbox{$>$}\;}
\def \lsim{\lower.8ex\hbox{$\sim$}\kern-.8em\raise.45ex\hbox{$<$}\;}

                      \def\ltsima{$\; \buildrel < \over \sim \;$}
                      \def\simlt{\lower.5ex\hbox{\ltsima}}
                      \def\rtsima{$\; \buildrel > \over \sim \;$}
                      \def\simrt{\lower.5ex\hbox{\rtsima}}

\title{
Non-Perturbative QCD Treatment of High-Energy Hadron-Hadron Scattering\\}

\author{H.G. Dosch}

\address{
Institut f\"ur Theoretische Physik der Universit\"at\\
Philosophenweg 16, D-69120 Heidelberg,FRG\\}

\author{Erasmo Ferreira}
\address{
Departamento de F\'\i sica, Pontif\'\i cia Universidade Cat\'olica\\
 Rio de Janeiro 22452 RJ, Brazil\\}

\author{A. Kr\"amer}
\address{ ~SAP, D-69190 Walldorf, FRG \thanks{~present address}\\}


\maketitle

\begin{abstract}

    Total  cross-sections and logarithmic slopes
of the elastic scattering cross-sections for different hadronic
processes are calculated in the framework of the model of the stochastic
vacuum. The relevant parameters of this model, a correlation length and
the gluon condensate, are determined from scattering data, and found to be
in very good agreement with values coming from completely different sources
of information. A parameter-free relation is given
between total cross-sections and slope parameters, which is shown
to be remarkably valid up to the highest energies for which data exist.
\end{abstract}
\bigskip
 PACS Numbers~:~12.38 Lg , 13.85 -t, 13.85 Dz , 13.85 Lg~.

\newpage
{\bf  1. INTRODUCTION }

It is certainly a great challenge to establish a microscopic description
of high-energy scattering in the framework of the field theory of strong
interactions, i.e. QCD. Sophisticated treatments of perturbation theory
\cite{RRA0} have lead to interesting results, which however are either
qualitative or not able to explain the most striking phenomena. There is
a rich amount of data on soft high-energy scattering , i.e. elastic
scattering at high energies and momentum transfers smaller than
the hadronic scale ($\approx$ 1 GeV). The newest and most precise data
come from proton-antiproton scattering \cite{RRA1}, extending up to center of
mass energies $\sqrt{s} = 1,800$ GeV. There are older data in the $pp$,$\pi p$
and $Kp$ channels  \cite{RRA2,RRA3}, and for other hadronic channels, such as
the $\Sigma p$ system\cite{RRA4}, they are still scarce. More soft-scattering
data are expected for the near future from Fermilab and LHC.

The energy dependence, in the full range of available data, is well
described in the Regge picture \cite{RRB1} . The total cross-sections
\cite{RRB2} increase with energy like $s^{0.0808}$, leading to a hypercritical
pomeron intercept. The variation of the slope of the elastic scattering
cross-sections is also well described in the Regge picture with a slope
of the pomeron trajectory $\alpha '(t)=0.25$ GeV$^{-2}$.

The value of about $2/3$ for the ratio of $\pi p$ to $p p$ (or $\bar p p $)
total cross-sections, as well as certain factorization properties, are
suggestive of an additive quark model \cite{RRB4}, in which the main
features of high-energy scattering can be described through quark-quark
scattering amplitudes. On the other hand, there is also a remarkable flavour
dependence of the cross-sections, which decrease with the increasing number
of strange quarks in the scattering channel. Such a feature is most naturally
explained in models in which
the cross-sections depend on the sizes of the hadrons, which is
also indicated by the dependence of the slopes of the elastic differential
cross-sections on the hadron sizes \cite{RRB5}.

 In this paper we evaluate
elastic scattering amplitudes of hadrons in the framework of the model
of the stochastic vacuum (MSV), originally developed in order to treat
non-perturbative effects in low-energy hadron physics \cite{RRC1,RRC2}. The
model therefore deals with parameters of non-perturbative QCD that play an
essential role both in hadron spectroscopy and in high-energy scattering.
Our treatment is based rather on loop-loop than on quark-quark scattering.
An important consequence is that the high-energy cross-sections depend
on the sizes of the hadrons, and this effect is due to the same mechanism
that leads to confinement.

The paper is organized as follows. In sec. 2 we give the theoretical
foundations of our model. In subsec. 2.1 we shortly recapitulate
essential aspects of the analysis of soft high-energy scattering
and in subsec. 2.2 some important features of the
model of the stochastic vacuum are explained. In sec. 3 we apply the MSV to
soft high-energy scattering and in sec. 4 we evaluate the formalism
derived in sec. 3 and find convenient numerical representations for the
results. In sec. 5 we discuss the choice of input parameters and compare our
theoretical results with experiment. In sec. 6 we present
concluding remarks. In the appendices we discuss some more technical points.

  \bigskip
{\bf  2. THEORETICAL BACKGROUND }

\noindent
2.1~{\bf Soft high-energy scattering in non-perturbative QCD}

 The first field theoretical approaches
to  soft high-energy scattering at small
momentum transfer were attempted by Amati, Fubini and Stanghelini \cite{RRD1}
in the framework of the multi-peripheral model and by Gell-Mann, Goldberger
and Low in a massive vector-exchange theory \cite {RRD2}. Since when it became
clear that QCD is the fundamental theory of the strong interactions,
many  efforts were made to explain the relevant features of soft
high-energy scattering, either by assuming genuine non-perturbative effects
\cite{RRD3} or by elaborate summation of perturbative
contributions \cite{RRA0} .

  The  success of methods, applied primarily in hadron spectroscopy,
taking into account  non-perturbative contributions as gluon-condensates
\cite {RRD4} stimulated Landshoff and Nachtmann \cite {RRD5} to apply
non-perturbative concepts also to soft high-energy
scattering. In an Abelian model they related elastic high-energy
scattering to the nontrivial structure of the QCD
vacuum, requiring, besides the (static) gluon condensate, a finite
correlation length for the slowly varying (non-perturbative) gluon
fields in the vacuum. In their Abelian model, this correlation leads
effectively to a non-perturbative gluon propagator. The main
consequences of the model were the
correct spin structure (vector-like exchange) of soft elastic
high-energy scattering
and quark additivity. The latter holds only if the above
mentioned correlation length of the gluon fields is small  compared
with the hadron radius. The energy dependence of the total cross section
comes out only approximately correct, namely as a
constant, instead of the experimentally observed
slow rise like $s^{0.0808}$. This model has been applied successfully also
to other channels \cite {RRD6}.

The structure of non-perturbative contributions to high energy scattering
was further investigated in a more general way by Nachtmann \cite {RRD7}, who
reduced the non-perturbative parts in soft high-energy hadron
scattering  to quark-quark (anti-quark) scattering, justifying this
reduction  in certain kinematical regions through considerations in the
femto-universe. Through the use of the eikonal
method, the high-energy scale was separated and the (non-perturbative) part
of the
quark-quark scattering amplitude that is due to exchange of states
with the vacuum quantum numbers could be reduced to an expression with
the  structure
\begin{eqnarray}
\big<q^{C'}(p_1)\ q^{D'}(p'_2)&|&q^C(p_1)\ q^D(p_2)\big>
\mathop{\longrightarrow}_{s\to\infty}  \nonumber \\
&&\bar u(p'_1)\ \gamma^\mu u(p_1)\
\bar u(p'_2)\ \gamma_\mu u(p_2)\ J_{qq}(q^2)\
\delta_{C'C}\delta_{D'D}~,
\label{2A1}
\end{eqnarray}
with small momentum transfer  $q=p'_1-p_1=p'_2-p_2$. The upper
indices $C, D$ denote
the colours of the quarks, and $u(p)$ is a Dirac spinor.

The transition from the quark-quark scattering amplitude  to the
observable hadron-hadron scattering amplitude was achieved through
current matrix elements occurring in deep inelastic scattering.

In this paper we widely follow the general analysis of
Nachtmann \cite{RRD7}, and evaluate quark-quark scattering amplitudes in a
specific non-perturbative model, namely that of a stochastic vacuum
with Gaussian fluctuations \cite{RRC1,RRC2} of the field strength.
There is however a very specific difference: whereas the original
treatment of Nachtmann is based on a reduction of  hadron-hadron
scattering to quark-quark scattering, our basic entities are
scattering amplitudes for Wilson loops in Minkowski
space-time. A definite advantage of our approach is the
gauge invariance of
these amplitudes in contradistinction to the quark-quark amplitudes.
 The loop amplitudes treated here can also be obtained in the
framework of Nachtmann  \cite{RRD7}, if one starts with hadron-hadron
rather than with quark-quark scattering matrix elements (O. Nachtmann,
private communication).

Another important difference must be remarked. Quark additivity appears
in a natural way in  the Landshoff-Nachtmann model
\cite {RRD5} and in the extended framework
of Nachtmann \cite{RRD7}. Here each quark interacts with the vacuum
field, and we may consider that this interaction defines a region with
the form of a tube around the quark path. Since the interaction with
another quark is due to the correlations of the fields in two such
tubes, the {\it effective} radius of a tube is actually determined
by the correlation length of the vacuum field. The $qq$ interactions
occurs only in the regions where two such tubes overlap.
If the separation of the quarks inside a hadron is large
compared with the correlation length $a$, the interaction
regions for the different pairs of quarks when two hadrons collide are
indeed well separated from each other, and quark additivity holds.
  However, such an argument is dangerous in a non-perturbative
treatment, since, for instance, it would
not lead to  the area law for the Wilson loop. Indeed, it is well
known that even a short range correlation of the fields can lead to
long range effects due to potentials, since the step from the fields to
the potentials is essentially non-local. A good example of this
phenomenon is the Bohm-Aharonov effect \cite{RRE1}, where  the phase of an
electron can be influenced by a magnetic field located far away. These long
range effects might
spoil quark additivity, and we show in sec.~2.2 that this is indeed
the case with the model of the stochastic vacuum : the same effect which
leads to confinement also leads to a violation of quark additivity. The
physical reason for that is easy to understand, since not only the quarks but
also the glue between them participate  essentially in the scattering
process. We return to this  point in technical detail in sec.~2.2.
\bigskip

\noindent{\bf 2.2~ The Model of the Stochastic Vacuum}

The model of the stochastic vacuum \cite{RRC1,RRC2} is based on the
idea that the low frequency contributions in the functional
integral can be taken into account by a simple stochastic process with
a converging cluster expansion \cite{RRE2}. This  assumption leads, in a
non-Abelian gauge theory, to linear confinement of static colour sources.

Let us phrase this idea in a somewhat more formal language.
Let ${\cal D}\tilde\phi(k)\ e^{-S[\tilde\phi]}$ be the functional
measure  of a quantum field theory where the fields to be integrated over are
expressed in momentum space. If the measure is split into low and
high frequency parts
\begin{equation}
{\cal D}\tilde\phi(k)=\prod_{|k|<\mu}{\cal D} \tilde \phi(k)
\prod_{|k| >\mu} {\cal D}\tilde\phi (k)~,
\label{2A2}
\end{equation}
the integration  over the high frequencies can be accounted for
perturbatively, in an asymptotically free theory.
All terms of higher than quadratic order in the fields
occurring in the exponential of the action $e^{-S[\bar\phi]}$ are
expanded in a power
series and, since only quadratic
terms are kept in the exponent, the remaining functional integrals
are Gaussian.

Little is known about the functional integration over the low frequencies.
In the QCD sum rule approach \cite{RRD4}, the contribution of this part
is taken into account by power corrections
proportional to specific non-perturbative vacuum expectation values
(condensates). A model which goes further is that of the stochastic
vacuum \cite{RRC1,RRC2} in its most restrictive form. Since we know that nature
has managed to regularize the infrared problems of perturbative theory
(after all, we do observe hadrons), we may assume that the integration
measure of the low frequency fields may be approximated by a simple
functional measure. The simplest ansatz is that of a Gaussian
integration measure, which is specified by a correlator
(corresponding to the propagator in perturbation theory). This
correlator is (apart from its specific form) determined by two scales:
the strength of the correlator and the correlation length.

Generically, we may write
\begin{equation}
\int\prod_{|k|<\mu}{\cal D}\tilde\phi(k) e^{-S}\ \phi(x_1)\ \phi(x_2) \equiv
  \langle \phi(x_1)\phi(x_2)\rangle_A   = G(x_1-x_2)~,
\label{2A3}
\end{equation}
and obtain all other Green's functions for a Gaussian process by
factorization. Note that we have liberally switched between the fields in
coordinate space, $\phi(x)$, and in momentum space, $\tilde\phi(k)$.
Eq.(\ref{2A3}) describes only the low internal frequencies part
of the correlator, and thus $G(x_1-x_2)$ is supposed to be regular for
$x_1\to x_2$. The singularities are due to perturbative terms.
As mentioned above, this simple model leads to confinement in a
non-Abelian gauge theory, and moreover the heavy quark potential
deduced from the correlation determining the Gaussian stochastic process
agrees very well with phenomenological  determinations \cite{RRE3,RRE4}.
Thus it is not unreasonable to apply the approximation by a Gaussian process
to other non-perturbative phenomena, as soft high-energy scattering.

First we discuss some basic properties of the two point correlator
defining the non-perturbative Gaussian process (i.e. the approximation for
the measure of functional integration over the low frequency fields).
 As always occurs with approximations, a
control of gauge invariance is essential. If that control is missing,
one is never sure whether the output of the calculation is only a gauge
artefact. So we do not deal with the correlator of the gauge potentials
$A^F_\mu(x)$, but rather of the field-strengths $F^F_{\mu\nu}$,
where the upper index $F$ is the colour index of the adjoint representation.
As usual, we introduce the Lie-algebra valued field quantities
\begin{equation}
{\bf A}_\mu(x)=\sum^{N^2_c-1}_{F=1} A^F_\mu(x)\ \tau_F\ ~ ,~ \
{\bf F}_{\mu\nu}(x)=\sum^{N_c^2-1}_{F=1}F^F_{\mu\nu}\ \tau_F \ ,
\label{2A5}
\end{equation}
and the covariant derivative
\begin{equation}
{\bf D}_\mu=\left({\bf 1}{\partial\over \partial
x_\mu}-ig{\bf A}_\mu\right) \ .
\label{2A6}
\end{equation}

In the expression above,
$\tau_F$ represents the $ N^2_c-1 $ generators of  the Lie algebra
of the gauge group $ SU(N_c) $. For $ N_c=3 $ they are, in the fundamental
representation, 1/2 times the Gell-Mann matrices  $ \lambda_F $.
If not stated otherwise, we always use the fundamental representation.
For some later discussions (see eq.(\ref{3A20}) below), it is convenient
to work with a general number of colours $N_c$.
In the first part of this section we work, as usual in the functional
approach, in an Euclidean space-time continuum, and therefore only lower
Lorentz indices are used.

In a non-Abelian gauge theory  the field strength tensor
${\bf F}_{\mu\nu}(x)$ does change under a local gauge transformation
\begin{equation}
{\bf F}_{\mu\nu}(x)\to {\bf U}(x)\  {\bf F}_{\mu\nu}(x)\ {\bf U}^{-1}(x)~,
\label{2A8}
\end{equation}
where ${\bf U}(x)$ is a local element of the gauge group $SU(N_c)$.

In order to give a well defined  meaning to a correlator, which is a
bilocal object, we parallel-transport the colour-content of all fields
to a single reference point $w$, i.e. we consider the parallel-transported
field strength tensor
\begin{equation}
{\bf F}_{\mu\nu}(x;w):=\phi^{-1}(x,w)\ {\bf F}_{\mu\nu} (x)\ \phi(x,w)~,
\label{2A9}
\end{equation}
where $\phi(x,w)$ is a non-Abelian Schwinger string from point $w$ to
 point $x$
\begin{equation}
\phi(x,w)={\bf P} \exp\left[-ig\int^1_0d\sigma ~ (x-w)_\mu
 {\bf A}_\mu(w+\sigma(x-w))\right] \ .
\label{2A10}
\end{equation}
${\bf P}$ denotes path ordering, which is necessary in order
to give
to the exponential a well defined meaning. In a non-perturbative way,
this ordering is defined, for any operator $\cal O$ through
\begin{equation}
{\bf P} \exp\left[i\int^1_0 {\cal O} (\sigma) \ d\sigma\right]=
\lim_{\Delta \sigma \to 0} \prod_k \exp
\left[i{\cal O}(\frac{\sigma_k+\sigma_{k-1}}{2})\cdot \Delta\sigma_k\right]~,
\label{2A11}
\end{equation}
with ~~$0<\sigma_1<\sigma_2\ldots <1~~,~ \
\Delta\sigma_k=\sigma_{k+1}-\sigma_k$~.

The field strength tensor in eq.(\ref{2A9}) transforms with the gauge
transformation at the fixed reference point $w$
\begin{equation}
{\bf F}_{\mu\nu}(x,w)\to {\bf U}(w)\ {\bf F}_{\mu\nu} (x;w)\ {\bf U}^{-1}
(w) \ .
\label{2A12}
\end{equation}

The correlator $\big<{\bf F}_{\mu\nu}(x,w)\
{\bf F}_{\delta\sigma}(y,w)\big>_A $,
  i.e. the vacuum expectation value with respect to the
low frequencies , is gauge-covariant because of eq.(\ref{2A12}); in
general, it may depend on the two coordinate differences $(x-w)$
and $(y-w)$. We
now make the crucial approximation that the correlator is independent of the
reference point $w$, and thus only depends on the difference $z=x-y$.
This approximation becomes  exact for $x\to y$, and is nearly true for $w$
fixed and large distances $z=x-y$. In this approximation, the most general
form of the correlator \cite{RRC2} is given by
\begin{eqnarray}
&\bigg<&g^2~F^C_{\mu\nu}(x,w)\ F^D_{\rho\sigma}(y,
w)\bigg>_A = {\delta^{CD}\over N^2_c-1}\ {1\over 12}\big<g^2~FF\big>
\nonumber \\
&\cdot&\bigg\{\kappa(\delta_{\mu\rho}\ \delta_{\nu\sigma}-
\delta_{\mu\sigma} \ \delta_{\nu\rho})\cdot D(z^2/a^2)   \nonumber  \\
 &+&(1-\kappa)\cdot {1\over 2}\Big[{\partial\over \partial
z_\mu}(z_\rho\delta_{\nu\sigma}-z_\sigma\delta_{\nu\rho})+
{\partial\over \partial z_\nu}(z_\sigma\delta_{\mu\rho}
-z_\rho \delta_{\mu\sigma})\bigg]
D_1(z^2/a^2)\bigg\}~.
\label{2A13}
\end{eqnarray}
Here $z=x-y$,
 $a$ is a characteristic correlation length,
$\big<g^2~FF\big>$ is the gluon condensate
\begin{equation}
\big<g^2~FF\big>=\big<g^2~F^C_{\mu\nu}(0)\ F^C_{\mu\nu}(0)\big>_A~,
\label{2A15}
\end{equation}
$N_c$ is the number of colours,~$C,D=1,\ldots,N_c^2-1$~, and
the factors in eq.(\ref{2A13}) are chosen in such a way that
\begin{equation}
D(0)=D_1(0)=1~.
\label{2A16}
\end{equation}

The two possible tensor structures are arranged in order that the
second term satisfies the homogeneous Maxwell equation, i.e.
\begin{equation}
{\partial\over \partial x_\beta}\
\varepsilon_{\alpha\beta\mu\nu}\bigg< g^2~F^C_{\mu\nu}(x,w)\
F^C_{\rho\sigma} (y,w)\bigg>_A= 0 \qquad {\rm for} \qquad \kappa=0 ~.
\label{2A17}
\end{equation}
Hence, in an Abelian gauge theory without monopoles, where the
homogeneous Maxwell equations must hold, only the second structure can
occur, i.e. we must there have $\kappa=0$.
However, in a non-Abelian theory there is no reason for $\kappa$ to be
zero.

With the form of eq.(\ref{2A13}) for the correlator, one
obtains \cite{RRC1,RRC2} the area
law for a Wilson loop with the string tension $\rho$, given by
\begin{equation}
\rho=\frac{\kappa \pi}{144}~\big<g^2~FF\big> a^2
\int^\infty_0 D(-u^2)\ du^2\ .
\label{2A18}
\end{equation}
Thus only the tensor structure proportional to $D$ leads to
confinement. This result has the very welcome consequence that
only in non-Abelian gauge theories the model of
the stochastic vacuum leads to confinement.  It also teaches us
that the correlator
$D$ is the part specific to non-Abelian gauge theories.

The correlator in eq.(\ref{2A13}) has been calculated on
the lattice \cite{RRF1},
and the results show unambiguously that $\kappa$ is different from
zero, as predicted by the model of the stochastic vacuum.
The ratio $\kappa/(1-\kappa)$ is rather large (about 3), so that
$D(z^2/a^2)$ is the dominant contribution. We return to this point
in more detail later.

\bigskip
{\bf 3. THE MODEL OF THE STOCHASTIC VACUUM IN HIGH-ENERGY SCATTERING}

The correlator in eq.(\ref{2A13}), specifying the Gaussian
   process that approximates
non-perturbative effects of QCD, is the starting point for our
evaluation of observables in soft high-energy scattering. In the
analysis mentioned in sec.~2, Nachtmann \cite{RRD7}  evaluated the
quark-quark scattering amplitude using the eikonal approximation for
the interaction of the quarks with the gluon field. In a first step,
we follow the same
approach, and consider the scattering amplitude of a single quark in a
given external colour potential ${\bf A}_\mu$. If the energy of the
quark is very high and the background field has only a limited
frequency range, the quark  moves on an approximately straigth
light-like line and the eikonal approximation can be applied. At the
end of this section we recall the condition for the validity of the eikonal
approximation.

 Along its path $\Gamma$, the quark picks up the
eikonal phase (which is here a unitary $N_c\times N_c$ matrix)
\begin{equation}
V={\bf P}\exp[-ig \int_\Gamma {\bf A}_\mu(z)\ dz^\mu]~.
\label{3A1}
\end{equation}
Here ${\bf A}_\mu$ is again the Lie-algebra valued potential and
${\bf P}$ denotes path ordering (see eq.(\ref{2A11})). The phase factor
for an antiquark is obtained  by complex conjugation.

{}From the scattering amplitudes for single quarks in the background
field,
we obtain the non-perturbative quark-quark scattering amplitude by
functional integration over the background field of the
product of the two scattering amplitudes. More specifically,
consider two quarks travelling along the light-like paths $\Gamma_1$ and
$\Gamma_2$ given by
\begin{equation}
\Gamma_1=(x^0,\vec b/2, x^3=x^0)~~~{\rm{and}}~~\ \Gamma_2= (x^0, -\vec b/2,
x^3=
-x^0) \ ,
\label{3A2}
\end{equation}
corresponding to quarks moving  with velocity of light in opposite
directions, with an impact vector $\vec b$ in the $x^1 x^2$-plane
(referred to in the following as the transverse plane). Let
$V_{1,2}(\pm\vec b/2)$ be the phases picked up by the quarks along these paths
\begin{equation}
V_{1,2}(\pm\vec b/2)={\bf P}\exp\left[-ig \int_{\Gamma_{1,2}} {\bf A}_\mu
(z)\ dz^\mu\right]~.
\label{3A3}
\end{equation}
Then the scattering amplitude for two quarks with momenta $p_1$, $p_2$
and colour indices $c_1$, $c_2$ leading to two quarks of
momenta $p_3$, $p_4$ and
colours  $c_3\ c_4$ is given by \cite{RRD7}
\begin{equation}
T_{c_3c_4;c_1c_2}(s,t)=\bar u (p_3)\ \gamma^\mu(p_1)\ \bar u(p_2)\
\gamma_\mu u(p_2)~{\cal V}~,
\label{3A4}
\end{equation}
where
\begin{equation}
{\cal V}=i\big<Z^{-2}_\psi\big>_A\bigg<\int d^2\vec b\ e^{-i\vec q \cdot
\vec b}\bigg\{\bigg[ V_1\Big(-{\vec b\over 2}\Big)-{\bf 1}\bigg]_{c_3c_1}
\bigg[V_2 \Big(+{\vec b\over 2}\Big)-{\bf 1}\bigg]
_{c_4c_2}\bigg\}\bigg>_A~.
\label{3A5}
\end{equation}
Here $\big<\ \big>_A$ denotes functional integration over the
background field; $\vec q$ is the momentum transfer $(p_1-p_3)$ projected
on the transverse plane. Of course the approximation makes sense only
if $|\vec q|\ll|\vec p|$. The quantity $Z_\psi $ is the fermion wave-function
renormalization constant in the eikonal approximation, given by \cite{RRD7}
\begin{equation}
Z_\psi[A]={1\over N_c}{\rm tr}~[V_1(0)]={1\over N_c} {\rm tr}~[V_2(0)]~.
\label{3A6}
\end{equation}
The subtraction of the unit operator from the phase-matrices $V$ is due
to the transition from the $S$ to the $T$-operator.

In the limit of high energies we have  helicity conservation
\begin{equation}
\bar u(p_3)\ \gamma^\mu u(p_1)\ \bar u(p_4)\ \gamma_\mu u(p_2)
\mathop{\longrightarrow}_{s \to \infty} 2s\delta_{\lambda_1\lambda_3}
\delta_{\lambda_4 \lambda_2} \ ,
\label{3A7}
\end{equation}
where $\lambda_i$ are the helicities of the quarks and $s=(p_1+p_2)^2$.
In the following we can thus ignore the spin degrees of freedom.

The scattering amplitude in eq.(\ref{3A4}) is explicitly gauge dependent and
the
cautioning remarks made in the last section apply here. But we know that, in
meson-meson scattering, for each quark there is an antiquark
moving on a nearly parallel line. Furthermore, the meson
must be a colour singlet state under local gauge transformations. To
construct such a colourless state we have to parallel-transport the
colour content from the quark to the antiquark (or vice versa) in
the same way as discussed in sec.~2 for the field-strength
tensor. Since this parallel-transport of the colours is made by a
Schwinger string $\phi(x_q, x_{\bar q})$ (see eq.(\ref{2A10})), we
obtain for the
meson a Wilson loop whose light-like sides are formed by the quark and
antiquark paths, and front ends by the Schwinger strings (see
fig.~3.1). The direction of the path of an antiquark is
effectively the opposite of that of a quark, so that  the loop has a
well defined internal direction. The resulting loop-loop amplitude is
now specified, not only by the impact parameter, but also by the
transverse extension vectors.


We thus introduce
the loop-loop scattering amplitude
\begin{eqnarray}
J(\vec b,\vec R_1,\vec R_2)&&= \bigg[ \big< {1\over N_c}
{\rm tr}~W_1(0,\vec R_1)\big>_A\big<{1\over N_c}{\rm tr}~W_2(0, \vec
R_2)\big>_A \bigg]^{-1}  \nonumber     \\
&&\bigg<{1\over N_c}{\rm tr}\bigg[W_2\Big(-{\vec b\over 2},\vec R_2\Big)
-{\bf 1}\bigg] {1\over N_c}
{\rm tr}\bigg[W_1\Big(-{\vec b\over 2},\vec R_1\Big)
-{\bf 1}\bigg]\bigg>_A~,
\label{3A8}
\end{eqnarray}
where $W_1(-{\vec b}/ 2,\vec R_1)$ is the Wilson loop
\begin{equation}
W_1\left(-{\vec b\over 2},\vec R_1\right)={\bf P}\exp\left[ -ig
\oint_{\partial S_1}{\bf A}_\mu(z)\ dz^\mu\right] \ .
\label{3A9}
\end{equation}
The closed loop $\partial S_1$ is a rectangle whose long sides
are formed by the quark path $\Gamma_1^q=(x_0,\vec b/2+\vec R_1,x_3=x_0)$
and the antiquark path  $\Gamma_1^{\bar q}=(x_0,\vec b/2-\vec R_1,x_3=x_0)$ and
whose front sides are formed by lines from $(T,\vec b/2+\vec R_1,T)$  to
 $(T,\vec b/2-\vec R_1,T)$ for large positive and negative $T$ (we will
then take the limit $T\to \infty$).~
$W_2(\vec b/2,\vec R_2)$ is  constructed analogously. The first factor
in eq.(\ref{3A8}) is the loop renormalization constant that replaces the quark
field renormalization in eq.(\ref{3A5}).


Our next aim is to perform the functional integration over $A$ by
applying the model of the stochastic vacuum discussed in the preceding
section. Since the correlator is given in terms of the
parallel-transported field tensor  $ {\bf F}_{\mu\nu}(x,w)$, we have
first to transform the line integrals $\int {\bf A}_\mu dz^\mu$ through
integrals over the field tensor. This is done with the help of the non-Abelian
Stokes-theorem \cite{RRF2}. By deforming the path as indicated in fig.~3.2, we
can express the line integral $\oint_{\partial S}{\bf A}_\mu(z)\ dz^\mu$,
where the closed path $\partial S$ goes from $w$  to $x_{(1)}$ then
to $x_{(2)}$ and back to $w$, into the surface integral
$\int_S{\bf F}_{\mu\nu} (z,w)\ d\Sigma^{\mu\nu}(z)$, where
$d\Sigma^{\mu\nu}(z)$ , with $\mu < \nu$, is the element of the surface $S$
at point $z$.
Here we have used that, for a sufficiently small contour, we have
\begin{equation}
\oint_{\partial S} {\bf A}_\mu dz^\mu=\int
_S{\bf F}_{\mu\nu}d\Sigma^{\mu\nu} +{\cal O}(S^2)\ .
\label{3A10}
\end{equation}
In this way we obtain
\begin{equation}
{{\bf P}}\exp\int_{\partial S}-ig{\bf A}_\mu(z)\ dz^\mu= {{\bf P}}_S\exp \int_S
-ig{\bf F}_{\mu\nu} (z,w)\ d\Sigma^{\mu\nu}(z) \ ,
\label{3A11}
\end{equation}
where ${{\bf P}}_S$ now denotes surface ordering according to fig.~3.2.
Since the reference point $w$ in the correlator (\ref{2A13}) must be the same
for both fields,  we have to choose a common reference point for both
traces  in the product
\begin{equation}
\bigg<{\rm tr}\ \bigg[W_1\Big(-{\vec b\over 2},\vec R_1\Big)-{\bf 1} \bigg]
\ {\rm tr}\ \bigg [W_2\Big({\vec b\over 2},\vec R_2\Big)-{\bf 1} \bigg]
\bigg>_A~.
\label{3A12}
\end{equation}
We choose the point $w$ in the most symmetric way and then the surface
emerging from the loop $\partial S_1$ is formed by the sliding sides of a
pyramid with the loop $\partial S_1$ as basis and the point with
coordinates $w$ as apex; the same
holds for $\partial S_2$~(see fig.~3.3).


 Before its application to high-energy scattering, the model of the
stochastic vacuum must be translated from Euclidean space-time,
in which it is naturally formulated, to the Minkowski continuum.
Unfortunately we cannot go the other way and continue eq.(\ref{3A8}) to the
Euclidean continuum, which would be the safe way from the point of view
of the functional integration. However the Wilson loops occurring in
eq.(\ref{3A8})
have light-like sides which would shrink to a point if continued to
a space time continuum with Euclidean metric. We think that this is a
serious obstacle in all attempts to evaluate soft high-energy
amplitudes numerically on a lattice.

Since we cannot adapt the scattering amplitude to the Euclidean world,
we have to proceed the other way and adapt the model of the stochastic
vacuum to the Minkowski world. We are fully aware that this is by no
means a trivial step and, pending a better analytical understanding of
non-perturbative effects, we have to let the experiment decide on the
justification (Similar problems occur when one applies instanton
effects to high-energy scattering \cite{RRF3}). Thus we must translate the
correlation function
in eq.(\ref{2A13}) to the Minkowski world. This is obvious for the tensor
structure, where we just substitute $\delta_{\mu\nu}$ by $-g_{\mu\nu}$, etc,
but simple choices for the correlation functions like $\exp(-z^2/a^2)$
or $\exp(-\sqrt{z^2}/a)$ cannot be analytically continued in a meaningful
way. Therefore, we  must look for correlation functions $D(z^2/a^2)$ and
$D_1(z^2/a^2)$ that fall off for  negative $z^2$ values (corresponding
to Euclidean distances), and whose Fourier
transforms exist in Minkowski metric, since these will enter
essentially in the scattering amplitudes. An ansatz for the correlator
that fulfills this requirement can be written
in terms of the Fourier transforms
\begin{eqnarray}
\langle g^2\ F^C_{\mu\nu}(x,w)\ F^D_{\rho\sigma}(y,w)\rangle_A=
{\delta^{CD}\over N_c^2-1}\ {1\over 12}\langle g^2 FF\rangle \int
{d^4k\over (2\pi)^4}\ e^{-ik{(x-y)/a}} \nonumber \\
\Big\{ (g_{\mu\rho} g_{\nu\sigma}-
g_{\mu\sigma} g_{\nu \rho})\ \kappa i\widetilde D(k^2) \nonumber \\
+ (- g_{\nu\sigma}k_\mu k_\rho+g_{\nu\rho}k_\mu
k_\sigma-g_{\mu\rho}k_\nu k_\sigma +g_{\mu\sigma}k_\nu
k_\rho)(1-\kappa)\ i \frac{d\widetilde D_1(k^2)}{dk^2}\Big\} ~,
\label{3A13}
\end{eqnarray}
where
\begin{equation}
i\widetilde D(k^2)=\int d^4z D(z^2/a^2)\ e^{ikz/a}
\label{3A14}
\end{equation}
and
\begin{equation}
i \widetilde D_1(k^2)=\int d^4z D_1(z^2/a^2)\ e^{ikz/a}~.
\label{3A15}
\end{equation}

 After this choice is made, all functional integrations can be performed,
in principle. The quantities $W_{1,2}$ in eqs.(\ref{3A8}) and (\ref{3A9})
can be expressed
as surface integrals, according to eq.(\ref{3A11}).
The exponential being  expanded, the expectation value can be calculated
using eq.(\ref{3A11}) and assuming factorization in a Gaussian process
(see below).

 Before we enter into
details, we make two remarks which facilitate further calculations.
First we note that, since
\begin{equation}
\langle {\rm tr}~ W_1\rangle_A=\langle {\rm tr}~W_2\rangle_A= N_c~,
\label{3A16}
\end{equation}
the functional integral over the surface of one pyramid alone vanishes.
To see the formal  reason for eq.(\ref{3A16}), we remark that in the
evaluation of a single loop, say $\langle {\rm{tr}}~W_1\rangle_A$, only
the expectation value
$\langle e_+^\mu F_{\mu i}(x,w)e_+^\rho F_{\rho k}(y,w)\rangle_A $
occurs, where $e_+^\mu$ is the light-like
vector  $(1,0,0,-1)$, and $i, k$ ~(1 or 2) are indices of the transverse
plane.
These
correlators are zero by virtue of the tensor structures given in
eq.(\ref{3A13})  (note that $e_+^\mu e_{+\mu}=0$).
Therefore only the unit term contributes, leading to eq.(\ref{3A16}). This
is not
in contradiction to the area law in Euclidean space-time, since the
area of loops with light-like sides would be zero in Euclidean metric.

Due to this mechanism, the quantities
\begin{equation}
  Z_{\psi} =  \big< {1\over N_c}{\rm tr}~W_1(0,\vec R_i)\big>_A
\label{3A16A}
\end{equation}
that enter in the first factor of the expression in eq.(\ref{3A8})
are equal to one (no loop renormalization). In the same way only mixed
terms, from different pyramids, contribute for the expectation value in
eq.(\ref{3A8}), since the correlation functions arising from the expansion
of ${\rm tr\ } W_1(-\vec b/2,\vec R_1)$ alone, or equivalently
of ${\rm tr}~W_2(\vec b/2,\vec R_2)$, contain only field projections
$e^\mu_{-} F_{\mu i}$~ or~ $ e^\mu_{+}F_{\mu i}$~ respectively.

We next expand the exponentials $W_i$ in eq.(\ref{3A8}). Since
in the expansion of the trace of the exponential at
least two terms are necessary $({\rm tr}~\tau_A=0)$, and because of
eq.(\ref{3A16}) the lowest order
contribution to the loop-loop scattering amplitude is given by
\begin{eqnarray}
J(\vec b,\vec R_1,\vec R_2)=-(-ig)^4\Big({1\over
2!}\Big)^2 \ {\rm tr}~[\tau_{C_1}\tau_{C_2}]\ {\rm tr}~[\tau_{D_1}
\tau_{D_2}] \cdot       \nonumber \\
\int_{S_1}\prod^2_{i=1} d \Sigma^{\mu_i\nu_i}(x_i)
\int_{S_2}\prod^2_{j=1} d\Sigma^{\rho_j\sigma_j}(y_j)  \nonumber \\
{1\over N^2_c}\bigg\langle F^{C_1}_{\mu_1\nu_1}(x_1,w)\ F^{C_2}_{\mu_2
\nu_2} (x_2,w)\ F^{D_1}_{\rho_1\sigma_1}(y_1, w)\
F^{D_2}_{\rho_2\sigma_2} (y_2,w)\bigg\rangle_A~ \nonumber \\
           +{\rm higher~correlators}~.
\label{3A17}
\end{eqnarray}

We next apply the factorization hypothesis
\begin{eqnarray}
 \big<F^{C_1}F^{C_2}F^{D_1}F^{D_2}\big> &&=
 \big<F^{C_1}F^{C_2}\big>\big<F^{D_1}F^{D_2}\big>  \nonumber \\
 &&+\big<F^{C_1}F^{D_1}\big>\big<F^{C_2}F^{D_2}\big>
 +\big<F^{C_1}F^{D_2}\big>\big<F^{C_2}F^{D_1}\big>~,
\label{3A18}
\end{eqnarray}
where the arguments and the Lorentz indices of $F^{C_i},~ F^{D_i} $ are the
same as in eq.(\ref{3A17}).

  We have checked that the higher order terms are indeed small as compared
to the leading term, and therefore we neglect them in the following. In
this way the surface ordering indicated in eq.(\ref{3A11}) becomes irrelevant.

 It is convenient to introduce the eikonal function $\chi$
\begin{equation}
\chi(\vec b,\vec R_1,\vec R_2)=(-ig)^2\int_{S_1} d\Sigma^{\rho\mu} (x)\
\int_{S_2}d\Sigma^{\sigma\nu}(y)\ \langle F^C_{\rho\mu}(x,w)\
F^C_{\sigma\nu}(y,w)\rangle_A \ .
\label{3A19}
\end{equation}
Then the loop-loop amplitude $J(\vec b,\vec R_1,\vec R_2)$ is given to
the lowest order in the correlator by
\begin{equation}
J(\vec b,\vec R_1,\vec R_2)=-{1\over N^2_c}\ {1\over 8}\ {1\over
(N^2_c-1)} \bigg[\chi(\vec b,\vec R_1,\vec R_2)\bigg]^2~.
\label{3A20}
\end{equation}
We notice the presence of the colour suppression factor ${1/(N^2_c-1)}$
that always occurs
in interactions between colourless objects. The eikonal function is
determined by the geometry and by the correlator (\ref{3A13}).

For the Fourier transform  $\widetilde D(k^2)$ of the scalar
correlation function  $D$,
that enters in eq.(\ref{3A13}), we introduce an ansatz which
fulfills the requirements made above, and write
\begin{equation}
\widetilde D(k^2)= {-6 i A_nk^2\over{ (k^2-1/\lambda^2_n)^n}}\
{1\over \lambda^{2n-6} _n} ~\ , \ n\ge 4\ ,
\label{3A21}
\end{equation}
 with
\begin{equation}
D(z^2/a^2)=\int \widetilde D(k^2)\ e^{-ik{z/a}} {d^4k\over
(2\pi)^4}\ .
\label{3A22}
\end{equation}
These functions are discussed in Appendix~1.
The constants $A_n$ and $\lambda_n$ are fixed by the
normalization conditions
\begin{equation}
D(0)=1 \qquad \hbox{and}\qquad \int^\infty_0 du\ D(-u^2)=1\ .
\label{3A23}
\end{equation}
The second of these conditions allows the identification of $a$ as
a correlation length.

 The string tension for the correlator of the form (\ref{3A21}) can
be  obtained \cite{RRC2} from eq.(\ref{2A18}). With
 our choice for the correlator, we have \cite{RRG2}
\begin{equation}
\rho=\kappa\langle g^2~FF\rangle a^2\,{2\over 81}
(n-3) \left[{\Gamma(n-3)\over \Gamma(n-5/2)}\right]^2 ~.
\label{3A24}
\end{equation}

  The scalar function $D_1$ is completely independent from $D$, and may
have different values for the parameters $a$ and $n$. Lattice calculations
\cite{RRF1} show however that the forms of
$D$ and $D_1$ in the Euclidean region at large distances are similar,
with $D$ about 3 times larger than $D_1$. We show in Appendix~2 that
even with equal weigths for the two functions, the contribution of $D$
to high-energy soft scattering is by far the dominant one, so that we
can safely concentrate on this function.

Since analytic calculations are most easily done for
the case $n\to\infty$, previous calculations
have been made with that choice \cite{RRG2,RRG1A,RRG1B,RRG3},
but for $n=4$ there is good agreement with the form
(exponentially decreasing at large distances) indicated by
lattice
calculations in the Euclidean region \cite{RRF1}; we  therefore  adopt
this form in the present work.

Up to now we have only considered loop-loop scattering amplitudes. It
would be highly desirable to have a formalism relating
 these fundamental field theoretical entities to observables,
corresponding to the operator product expansion for quark (and gluon)
amplitudes. In the absence of such a formalism, we have to rely
on a rather simple minded quark model.

In a relativistic quark model, the distribution of the quarks
is described by the transverse momentum $k_\perp$ and the
fraction $x$ of the longitudinal momentum carried by the
quark. Since our amplitude is independent of the momentum of the
quarks (as long as the energy is high enough to ensure light-like paths), we
may neglect the dependence on $x$, and only consider the
transverse dependence. This transverse dependence is given by the Fourier
transform of the transverse wave-function, which determines the width
$2|\vec R|$ of the Wilson loops. We thus obtain our hadron-hadron (here
still meson-meson) scattering amplitude by smearing over the values of
$\vec R_1$ and $\vec R_2$ in eq.(\ref{3A17}) with transverse wave-functions
$\psi(\vec R)$. This leads to the meson-meson scattering amplitude
\begin{equation}
J_{MM'}(\vec b)=\int d^2\vec R_1~\int d^2\vec R_2~J(\vec b,\vec R_1, \vec R_2)
{}~\vert \psi_M(\vec R_1)\vert^2~\vert \psi_{M'}(\vec R_2)\vert^2~ .
\label{3A26}
\end{equation}
The choice of the transverse wave-functions $\psi_M(\vec R)$ will be
discussed in the next section.


For the treatment of the baryons we restrict ourselves to $N_c=3$. We
adopt two pictures: a genuine
three-body configuration, and a diquark picture. In the latter the baryon is
described exactly as a meson, where the diquark replaces the antiquark.
In the three body picture the baryon is described as shown
in fig.~3.4. There are three quark paths leading from $x_(i)$ to $x'_(i)$,
$i=1,3$. The coordinates $x_h$ and $x'_h$ refer to the central point of
the baryon. The paths from $x_h$ to $x_{(i)}$ and $x'_h$ to $x'_{(i)}$
respectively must ensure that the baryon is a colour singlet under local
gauge transformations. This is done by parallel-transporting the
colour from the quark positions $x_(i)$ to $x_h$, and coupling the colours
antisymmetrically in the form
\begin{equation}
{1\over \sqrt6}\,\varepsilon_{abc}\ \phi_{aa'}(x_h,x_{(1)})\
\phi_{bb'} (x_h,x_{(2)})\ \phi_{cc'}(x_h,x_{(3)})~,
\label{3A27}
\end{equation}
where $\phi$'s  are the Schwinger strings, eq.(\ref{2A10}). An analogous factor
occurs at the end (primed coordinates), so that the baryon is described
by the product of paths
\begin{equation}
{1\over 6}\,\varepsilon_{abc}\ \phi_{ad}[\Gamma_1]\
\phi_{be}[\Gamma_2] \ \phi_{cf}[\Gamma_3]\ \varepsilon_{def}~,
\label{3A28}
\end{equation}
where the path $\Gamma_i$ leads from $x_h$ over $x_{(i)}$ and $x'_{(i)}$ to
$x'_h$. Let $\Gamma_0$ be the path leading from $x'_h$ to $x_h$. Since
$\phi[\Gamma_0] \in SU(3)$, we have
\begin{equation}
\varepsilon_{def}=\varepsilon_{a'b'c'}\ \phi_{da'}[\Gamma_0] \
\phi_{eb'} [\Gamma_0]\ \phi_{fc'}[\Gamma_0]~.
\label{3A29}
\end{equation}
Inserting this expression into eq.(\ref{3A28}), we obtain that
the baryon is represented by
the product of Wilson loops  (without traces)
\begin{equation}
{1\over 6}\ \varepsilon_{abc} W_{aa'}[\partial S_1]\ W_{bb'}
[\partial S_2]\ W_{cc'} [\partial S_3] \ \varepsilon_{a'b'c'} \ ,
\label{3A30}
\end{equation}
where $W_{aa'}[\delta S_1]=\phi_{ad}[\Gamma_1]\ \phi_{da'} [\Gamma_0]$
is the loop from $x_h$ to  $x_1,x'_1,x'_h$ and back to $x_h$. Thus for a
baryon the factor $(1/ N_c){\rm tr}\ [W_1(-\vec b/2,\vec
R_1)-{\bf 1}]$ in eq.(\ref{3A8}) has to be replaced by
\begin{equation}
{1\over 36}\,\varepsilon_{abc}\varepsilon_{a'b'c'}\left\{ W_{a'a}
(-\vec b/2,\vec R_1) \ W_{b'b}(-\vec b/2,\vec R_2)\ W_{c'c}(-\vec b/2,
\vec R_3)-\delta_{aa'}\ \delta_{bb'}\ \delta_{cc'}\right\}~. \
\label{3A31}
\end{equation}
Here $\vec R_i$ is the vector extending from the middle line $\Gamma_0$
to the border of loop $i$. The impact parameter vector $\vec b$ is
taken with respect to the middle line $\Gamma_0$. The factor $1/36$ is
due to colour normalization. We discuss baryon transverse
wave-functions in the next section.

We conclude this section with some considerations concerning the
validity of the underlying quark model, reviewing  the analysis
made by Nachtmann \cite{RRD7}, who discusses
limits on the energy values $\sqrt{s}$ in order
to ensure the validity of his treatment. It is clear that the eikonal
approximation
 can be justified only for a sufficiently high energy.
A detailed analysis yields the condition
$\sqrt {s} \geq 2 \tau_0 h^{-2} $
where $h$ is a typical hadronic scale  ($h \sim 1$ GeV$^{-1}$) and
$\tau_0$ is the
observation time of the scattering process in the {\it femto universe}. There
is another scale $Q_0$ that indicates the separation between the
perturbative and the non-perturbative effects. The correlation length
and the wave-function depend on its value, since it can be viewed
as a renormalization scale for non-perturbative quantities. The value of
$Q_0$ should also be approximately equal to the hadronic scale
$h^{-1}$, and its relation with the energy and the observation time was
found \cite{RRD7} to be
\begin{equation}
Q_0^2 \approx \frac{\sqrt{s}}{2 \tau_0}~.
\label{3A32}
\end{equation}
{}From this relation Nachtmann also obtained a conservative {\em upper}
bound for the energy, which results from the requirement that the
observation time should be short enough so that the string between
the scattering quarks or antiquarks is not broken if the two particles
fly apart. The string breaks if the two quarks become so far apart that
the potential energy of the string reaches
the {\it proto-hadron} mass $m_c\approx 1.3$ GeV . Then the breaking
time $\tau_{cr} $ is determined by
\begin{equation}
\tau _{cr} = \frac {m_{c}}{ \rho}~,
\label{3A33}
\end{equation}
where $\rho$ is the string tension
$\rho \approx 0.18$ GeV$^{-2}$~.
The maximal observation time is given by $\tau_0=2 \tau_{cr}$.
In the present paper we consider only elastic scattering of loops, where no
string between
quarks of different hadrons is formed, hence we see no compelling reason
to apply the above limit to $\tau_0$, and there is no upper limit on
the energy for the applicability of the model from this consideration.
However, for definiteness, we take in the present work as a
reference energy the value $\sqrt{s}
= 20$ GeV, corresponding to the conservative upper bound  indicated by
Nachtmann \cite{RRD7}, and discuss other energies separately.

\bigskip
\vfill
{\bf 4. NUMERICAL EVALUATION AND PARAMETRIZATION OF THE RESULTS   }

     We now introduce the notation $\vec R(I,L)$, where the first index
(I=1,2) specifies the hadron, and the second specifies the particular
quark or antiquark in that hadron.
     We first evaluate the eikonal functions
$ \chi (\vec b,\vec R(1,1),\vec R(2,1))$ in eq.(\ref{3A19})
for the confining case, namely $\kappa =1$. The integration surfaces
$S_1$ and $S_2$ of the eikonal function
\begin{equation}
\chi (\vec b,\vec R(1,1),\vec R(2,1))
= (-ig)^2\int_{S_1}d\Sigma^{\mu\nu}(x)
     \int_{S_2}d\Sigma^{\rho\sigma}(x')~\langle F^C_{\mu\nu}(x,w)
     F^C_{\rho\sigma}(x',w)\rangle_A ~
\label {4A1}
\end{equation}
are represented in figs.~3.3 and 4.1. The first of these figures gives a
somewhat tilted three-dimensional view, while the second shows a projection
on the transverse plane. The vectors $\vec Q(K,L)$ in the transverse
plane connect the reference point $C$ (with coordinates $w$) to the
positions of the quarks and antiquarks of the loops 1 and 2. The quantity
$\psi(K,L)$ is the angle between $\vec Q(1,K)$ and $\vec Q(2,L)$.


    In the integrations indicated
in eq.(\ref{4A1}) we first note that the contributions
involving front planes ($CA_{11}A_{12}$, $CB_{11}B_{12}$, $CA_{21}A_{22}$  and
$CB_{21}B_{22}$) in fig.~3.3 vanish in the limit $ T\to \infty$, and
therefore we are left with four
remaining terms, the integrals over the products of the side planes of two
different pyramids. The four-dimensional integration over the two surfaces
can be finally reduced to a single integration \cite{RRG2}. The integrations
along the directions $x_+$ and $x_-$ can be performed, and result in
expressions involving two-dimensional inverse Fourier transforms of the
correlator $\widetilde D(k^2)$. A typical resulting contribution
that comes from the product of surfaces $(CA_{11}B_{11})$
and $(CA_{21}B_{21})$ in fig.~3.3 is
\begin{equation}
\int_0^1 d\alpha \int_0^1 d\beta~\cos\Psi(1,1)~{\cal{F}}_2^{(4)}
         (-\vert \alpha\vec Q(1,1)-\beta \vec Q(2,1)\vert^2)~,
\end{equation}
where ${\cal{F}}_2^{(4)}$ is the above mentioned two-dimensional Fourier
transform of the correlator with $n=4$ (see eq.(\ref{3A21})), defined as
 \begin{equation}
                    {\cal F}_{2}^{(4)}(-|\vec\xi|^2)
    = - \frac{1}{(2\pi)^2} \int{  {d^2\vec K_\bot}~
         \frac{6 A_4\lambda_4^{6}|\vec K_\bot |^2 }
         {(-\lambda_4^2|\vec K_\bot|^2-1)^4}
          ~\exp{({i\vec K_\bot}\cdot {\vec\xi})} }~,
\label{4A2}
\end{equation}
     where $\vec \xi$
      is a two-dimensional vector of the transverse plane.
This quantity, which is evaluated in Appendix 1, can be can be written
in the convenient  form
  \begin{equation}
               {\cal F}_{2}^{(4)}(-|\vec\xi|^2)
 = -\frac{32}{9\pi}\Delta_2
             [(\rho|\vec\xi|)^{3} K_{3}(\rho|\vec\xi|)]~,
\label{4A3}
\end{equation}
  where
\begin{equation}
              \rho= \frac{3\pi}{8}~,
\label{4A3A}
\end{equation}
 $K_3$ is a modified Bessel function,  and $\Delta_2$ is the Laplacian in
two dimensions.

             Taking advantage of the Laplacian form, we can apply
Gauss' theorem in two dimensions and eliminate one further
integration \cite{RRG2}.

              It is useful to introduce a reduced eikonal function and
a reduced loop-loop scattering amplitude through
\begin{equation}
    \widetilde \chi(\vec b, \vec R_1, \vec R_2) \equiv
    \frac{12}{\kappa \langle g^2 FF \rangle} \chi (\vec b, \vec R_1, \vec R_2)
\label{4A4}
\end{equation}
and
\begin{eqnarray}
    \widetilde J_{MM'}(\vec b, \vec R_1, \vec R_2) \equiv
\frac{1}{\big[\kappa \langle g^2 FF \rangle\big]^2}
 J_{MM'}(\vec b, \vec R_1, \vec R_2)=
-\frac{[ \widetilde \chi(\vec b, \vec R_1, \vec R_2)  ]^2 }
{144 \cdot 8 \cdot  N_c^2(N_c^2-1)}~.
\label{4A5}
\end{eqnarray}
(see eq.(\ref{3A20})). We then obtain

\begin{eqnarray}
  \widetilde \chi ( \vec b,\vec {R}(1,1),\vec {R}(2,1))=
          -\cos \psi (1,1)~I[Q(1,1),Q(2,1),\psi (1,1)] ~~ \nonumber \\
          -\cos \psi (2,2)~I[Q(1,2),Q(2,2),\psi (2,2)] ~~ \nonumber \\
          +\cos \psi (1,2)~I[Q(1,1),Q(2,2),\psi (1,2)] ~~ \nonumber \\
          +\cos \psi (2,1)~I[Q(1,2),Q(2,1),\psi (2,1)] ~,
 \label{4A6}
\end{eqnarray}
     where the quantities {\it {I}} are given by

\begin{eqnarray}
  I[Q(1,K),Q(2,L),\psi (K,L)] = {\frac {32}{9\pi}}
                         \bigg(\frac{3\pi}{8}\bigg)^2
\quad\quad\quad\quad\quad\quad\quad\quad\quad \quad \nonumber \\
\quad\quad  \times \{  Q(1,K)
   \int_{0}^{Q(2,L)} [Q(1,K)^2+x^2-2 x Q(1,K) \cos{\psi(K,L)} ]
                                     \nonumber  \\
K_{2}\bigg[\frac{3\pi}{8}\sqrt{Q(1,K)^2+x^2
    -2 x Q(1,K) \cos{\psi(K,L)}}\bigg]
                                ~dx ~~~                 \nonumber \\
                                                        \nonumber \\
    + Q(2,L)
   \int _{0}^{Q(1,K)}  [Q(2,L)^2+x^2-2 x Q(2,L) \cos{\psi(K,L)} ]
                                     \nonumber  \\
K_{2}\bigg[\frac{3\pi}{8}\sqrt{Q(2,L)^2+x^2
    -2 x Q(2,L) \cos{\psi(K,L)}}\bigg]
                                ~dx               \}~,
\label{4A7}
\end{eqnarray}
with  $Q(K,L)=\vert\vec Q(K,L)\vert$.

 From the eikonal function $\widetilde \chi$ we contruct the loop-loop
amplitude $ \widetilde J^{MM'}(\vec b, \vec R_1, \vec R_2)$
following eq.(\ref{3A20}), where $\vec R_1$ and $\vec R_2$ are shorthand
notation for $\vec R(1,1)$ and $\vec R(2,1)$ respectively.
The meson-meson scattering amplitude is then constructed by averaging
over the transverse wave-functions, according to eq.(\ref{3A26}).

         These results apply equally well to meson-baryon and baryon-baryon
scattering if the baryon is represented as a meson-like structure in
the quark-diquark picture.

         The evaluation of the eikonal function and of the observables
for the non-confining correlator follows the same lines. Since this part
of the correlator is not going to be used in the phenomenological analysis,
we only present in the Appendix~2 the comparison between some
corresponding quantities for the two cases,
in order to exhibit their large differences.


        In order to treat the baryon as a genuine three-body configuration,
we have to start from the expression given in eq.(\ref{3A30}). The projection
of the loops for meson-baryon scattering is shown in fig.~4.2. In this case
the relation between the scattering
amplitude $J$ and the eikonal functions $\chi$ is more
complicated than eq.(\ref{3A20}). Let us now define  $\widetilde\chi$  by

\begin{eqnarray}
    A_K \equiv \widetilde \chi (\vec b,\vec {R}(1,K),\vec {R}(2,1))   =
       -\cos \psi (K,1)~I[Q(1,K),Q(2,1),\psi (K,1)]~~  \nonumber \\
          +\cos \psi (K,2)~I[Q(1,K),Q(2,2),\psi (K,2)] ~,
\label{4A8}
\end{eqnarray}
with the functions {\it {I}} given  by eq. (\ref {4A7}).
  The index $K=1,2,3$ refers
to the three quarks in the baryon loop, which is here taken as hadron
number 1. As before, in the meson loop (hadron number 2) the quark is
labelled 1, the antiquark is labelled 2. The relevant quantity for the
transition probability is then
\begin{eqnarray}
  \widetilde J_{BM}[\vec b, \vec R(1,1), \vec R(1,2), \vec R(1,3), \vec R(2,1)]
     =      \nonumber \\
\frac{-1}{144\cdot 8 \cdot N_C^2(N_C^2-1)}
   (A_1^2+A_2^2+A_3^2-A_2A_3-A_1A_3-A_1A_2) ~.
\label{4A9}
\end{eqnarray}

   For the  scattering between two baryons in the three-body picture
we define $\widetilde \chi$ by

\begin{equation}
   \widetilde \chi (\vec b,\vec {R}(1,K),\vec {R}(2,L) ) \equiv A_{KL} =
          -\cos \psi (K,L)~I[Q(1,K),Q(2,L),\psi (K,L)]  ~,
\label{4A10}
\end{equation}
which obviously represents the contribution due to the interaction
between the quark loop $K$ in baryon 1 and the quark loop $L$ in
baryon 2. Then the hadron-hadron scattering amplitude is given by
\begin{eqnarray}
  \widetilde J_{BB}[\vec b, \vec R(1,1), \vec R(1,2), \vec R(1,3),
 \vec R(2,1), \vec R(2,2), \vec R(2,3)]=
       \frac{-1}{144\cdot 8 \cdot N_C^2(N_C^2-1)} \nonumber \\
        \bigg[
    A_{11}^2+A_{12}^2+A_{13}^2+A_{21}^2+A_{22}^2+A_{23}^2
                              +A_{31}^2+A_{32}^2+A_{33}^2~~~\nonumber \\
   -A_{12}A_{13}-A_{22}A_{23}-A_{32}A_{33}-A_{11}A_{13}-A_{21}A_{23}
   -A_{31}A_{33}                                      ~~~  \nonumber \\
   -A_{11}A_{12}-A_{21}A_{22}-A_{31}A_{32}
   -A_{21}A_{31}-A_{22}A_{32}-A_{23}A_{33}            ~~~  \nonumber \\
   -A_{11}A_{31}-A_{12}A_{32}
   -A_{13}A_{33}-A_{11}A_{21}-A_{12}A_{22}-A_{13}A_{23} ~~~\nonumber \\
  + \frac {1}{2} \bigg(
    A_{22}A_{33}+A_{23}A_{32}+A_{21}A_{33}+A_{23}A_{31}+A_{21}A_{32}
   +A_{22}A_{31}                                       ~~~ \nonumber \\
   +A_{12}A_{33}+A_{13}A_{32}+A_{11}A_{33}
   +A_{13}A_{31}+A_{11}A_{32}+A_{12}A_{31}            ~~~  \nonumber \\
   +A_{12}A_{23}+A_{13}A_{22}
   +A_{11}A_{23}+A_{13}A_{21}+A_{11}A_{22}+A_{12}A_{21} \bigg)\bigg] ~.
\label{4A11}
\end{eqnarray}

          In the simplest case we place the quarks around the central point
of the baryon in such a way that the vectors $\vec R(1,K)$ form angles
of 120 degrees (this configuration minimizes the string tension), and
also choose the distances to the baryon center to be all equal, i.e.
$\vert\vec R(1,1)\vert=\vert\vec R(1,2)\vert=\vert\vec R(1,3)\vert$. In this
way, to form the amplitude, the configuration
of the loops forming a baryon is specified by only one transverse vector,
say $\vec R(1,1)$ of quark 1.

           For the hadron transverse wave-function we make the ansatz
   of the simple  Gaussian form
\begin{equation}
     \psi _{H} (R)=  \sqrt {2/\pi}\frac {1}{S_H} \exp{(-R^{2}/S_H^{2})}~,
\label{4A12}
\end{equation}
     where $S_H$  is a parameter for the hadron size.

      Analogously to eq.(\ref{3A26}),
          we write the reduced hadron-hadron amplitude as an average
     over the hadronic wave-functions
\begin{equation}
\widehat J_{H_1 H_2}(\vec b,S_1,S_2)=
       \int d^{2}\vec R_{1}\int d^{2}\vec R_{2}~
  \widetilde J_{H_1 H_2}(\vec b,\vec R_1,\vec R_2)~
           {\vert\psi_1(\vec R_1)\vert}^2
           {\vert\psi_2(\vec R_2)\vert}^2~,
\label{4A13}
\end{equation}
    which is a dimensionless quantity.

     The hadron-hadron scattering amplitude in the eikonal approach
is then given by
\begin{equation}
T_{H_1 H_2} = i 2 s [\kappa\langle g^2 FF\rangle]^2 a^{10}\int d^2 \vec b~
          \exp{(i \vec q\cdot \vec b )}~\widehat J_{H_1 H_2}
(\vec b,S_1,S_2) ~,
\label{4A14}
\end{equation}
  where the impact parameter vector  $\vec b$  is in units of the
     correlation length $a$, and $ \vec q $ is the momentum
     transfer projected on the transverse plane (in units of $1/a$,
     so that the momentum transfer squared is $t=-|\vec q|^2/a^2$). The
 eikonal approach requires large $s$ and  $|t|<<s$.

         We have verified that the contributions of next order in
$\langle g^2 FF \rangle$ are small \cite{RRG2}, which is a consequence
of the presence of the colour factor in eq.(\ref{3A20}).

              Our normalization
 for $T_{H_1 H_2}$ is such that the total
 cross-section is obtained through the optical theorem by
\begin{equation}
            \sigma^T  = \frac{1}{s}~\hbox{\rm Im}~T_{H_1 H_2} ~,
\label{4A15}
\end{equation}
  and the differential cross-section is given by
\begin{equation}
    \frac {d\sigma^{e\ell}}{dt} =
             \frac {1}{16\pi s^2}~\vert T_{H_1 H_2}\vert^2  ~.
\label{4A16}
\end{equation}

    For short, from now on we write $J(b)$  to represent
$\widehat J(\vec b,S_1,S_2)$.

     The shapes of $J(b)$ for the three cases of hadronic scattering
     are shown in fig.~4.3, against the impact parameter $b$ (in units
     of the characteristic length $a$). In all the three curves
     represented in the figure, we
     have used $ S/a = 4$ for both interacting hadrons.
     The label $M$ means meson-like structure, while
     $B$ means a three-body configuration.


        In all cases, $ J(b/a)$ can be written, in very good approximation,
     as a function of the form
\begin{equation}
 J(b)=J(0)~\frac{P_1+P_2 (b/a)^2}{P_1+(b/a)^2}~\exp{(-P_3 (b/a)^2)} ~,
\label{4A17}
\end{equation}
 where $P_1$, $ P_2$, $ P_3$ are parameters, determined by fitting the
      exact (numerically obtained) values of $J(b/a)$.

        Let us define the dimensionless quantities (as before, with $b$
in units of the correlation length $a$)
\begin{equation}
              I_k = \int d^2\vec b~b^k~J(b) ~~,~k=0, 1, 2, ...
\label{4A18}
\end{equation}
which depend only on $ S/a$, and the Fourier-Bessel transform
\begin{equation}
              I(t) = \int d^2\vec b~ J_0 (b a \sqrt{|t|})~J(b) ~,
\label{4A19}
\end{equation}
where $J_0(b a \sqrt{|t|})$ is the zeroth--order Bessel function.
Thus $$T_{H_1 H_2} = i s[\kappa\langle g^2 FF\rangle]^2 a^{10} I(t)~. $$

 Since  $J(b)$ is real, the total cross section $\sigma^T$, the
differential elastic cross-section and the
slope parameter(slope at $t=0$)
\begin{equation}
   B = \frac {d}{dt} \biggl( \ln \frac {d\sigma^{e\ell}}{dt} \biggr)
   \bigg\vert_{t=0} ~,
\label{4A20}
\end{equation}
are given by
\begin{equation}
            \sigma^T= I_0~\kappa\langle g^2 FF \rangle^2 a^{10} ~,
\label {4A21}
\end{equation}
\begin{equation}
\frac {d\sigma}{dt} = \frac {1}{16\pi} I(t)^2~
                     [\kappa\langle g^2 FF \rangle]^4 a^{20} ~,
\label{4A22}
\end{equation}
and
\begin{equation}
          B=\frac{1}{2}~ \frac{I_2}{I_0} ~a^2 ~ = K a^2.
\label{4A23}
\end{equation}
 We have here defined
\begin{equation}
         K= \frac{1}{2}~ \frac{I_2}{I_0} ~,
\label{4A24}
\end{equation}
which is a function of $S_1/a$ and $S_2/a$ only.

    We observe that in the lowest order of the correlator expansion used
here, the  slope parameter $B$ does not depend on the value of the gluon
condensate  $ \langle g^2 FF \rangle $.

    The curves for
 $I_0= \sigma^T /~[(\kappa g^2\langle FF \rangle)^2 a^{10}] $~
and for $ K= B/a^2 $ as functions of $S/a$
are shown respectively in figs.~4.4 and 4.5, for
the $BB$, $MM$ and $MB$  cases, with hadrons of same extensions,
$S_1=S_2=S$.

%

    In the interesting ranges  $1\leq S/a\leq3$ and $0.5\leq S_2/S_1 \leq1$,
parametrizations for the dependence of the total
cross-section and slope parameter on the hadron size parameters $S_i$, to
an accuracy better than $3 \%$ are
\begin{equation}
I_0= \alpha \bigg(\frac{S_1 S_2}{a^2}\bigg)^{\beta/2}
\label{4A25}
\end{equation}
and
\begin{equation}
B=1.558 a^2+(\gamma/2) [S_1^2+S_2^2]~,
\label{4A26}
\end{equation}
where the value of the constants $\alpha, \beta, \gamma$ are given
in table 4.1~ for the three different cases .

    \vspace{1.5cm}
    Table 4.1~ Values of the parameters to be used in eqs.
(\ref{4A25}) and (\ref{4A26}) to determine total cross-sections and slopes.

\bigskip
\begin{center}
\begin{tabular}{|c|c|c|c|c|c|c|c|c|c|}
\hline
Case  & $\alpha$    &   $\beta$ &    $\gamma$ \\
\hline
$MM$  & 0.00626   & 3.090   &     0.366 \\
$BB$  & 0.00881   & 3.277   &     0.454 \\
$BM$  & 0.00682   & 3.135   &     0.348 \\
\hline
\end{tabular}
\end{center}
    \vspace{1.0cm}

     A more precise parametrization, accurate in ranges of $S_1/a$
extending from 0 to 5, is given by
\begin{equation}
I_0= \alpha \bigg(\frac{S_1}{a}\bigg)^{\beta}
\label{4A27}
\end{equation}
and
\begin{equation}
K=1.558+\gamma \bigg(\frac{S_1}{a}\bigg)^{\delta}~,
\label{4A28}
\end{equation}
where now the coefficients $\alpha$ and $\gamma$ depend on the
ratio $S_2/S_1$.
       The values of the parameters  $\alpha,\beta,\gamma,\delta$
are given in table 4.2, for values of the ratios $S_2/S_1$  which
we consider in the next section, where, besides $pp$ and $\bar p p$
 scattering, we also discuss
$p\pi$, $pK$ and $p\Sigma$ scattering, in which cases $S_2\not=S_1$.

\vspace{1.5cm}
    Table 4.2~ Values of the parameters for eqs. (\ref{4A27}) and
(\ref{4A28}), for several values of $S_2/S_1$ which are important to
represent the total cross-sections and slopes for different hadronic
systems.

\bigskip
\begin{center}
\begin{tabular}{|c|c|c|c|c|c|c|c|c|c|}
\hline
Hadrons&1 & 2 &  \multicolumn {2}{|c|}{ parameters for cross-sections } &
\multicolumn {2}{|c|}{ parameters for slopes }                  \\
\hline

     $  $ & $ $& $ $ &$\alpha $ & $\beta$ &
  $\gamma $& $\delta$  \\
\hline
$S_2=S_1$& M & M      & 0.006260 &  3.090  & 0.3616 &  2.023     \\
    & B & M        & 0.007846 &  3.135  & 0.4311 &  1.955           \\
   & B & B        & 0.008814 &  3.277  & 0.4891 &  1.892            \\
\hline

$S_2=0.94 S_1$& M & M      & 0.005610 &  3.090  & 0.3403 &  2.023     \\
   & B & B        & 0.008000 &  3.277  & 0.4636 &  1.892            \\
\hline

$S_2=0.77 S_1$ & M & M   & 0.004159 &  3.090  & 0.2908 &  2.023      \\
 &  B & M        & 0.004446 &  3.277  & 0.3577 &  1.955            \\
\hline
$S_2=0.67 S_1$ &M & M    & 0.003353 &  3.090  & 0.2672 &  2.023       \\
 &  B & M        & 0.003604 &  3.277  & 0.3330 &  1.955            \\
\hline
\end{tabular}
\end{center}
  \vspace{1.0cm}

      The t-dependence of the logarithmic slope of the differential
 cross-section is given by
\begin{eqnarray}
 B(t)&=& \frac {d
}{dt} \biggl( \ln \frac {d\sigma^{e\ell}}{dt} \biggr)
     = \frac{2}{I(t)} \frac{dI(t)}{dt}    \nonumber \\
&=&\frac{a^2}{I(t)} \int _{0}^{\infty} \frac{2\pi}{\sqrt{|t|}}
     J(b) b^2 J_1(b a \sqrt{|t|}) db ~.
\label{4A29}
\end{eqnarray}

  For small $t$ values, we obtain, expanding the Bessel function,
\begin{eqnarray}
 B(t)&\simeq&B(0) \bigg[1+\frac{1}{8}\bigg(\frac{2I_2}{I_0}-\frac{I_4}{I_2}
          \bigg) a^2 |t| \bigg]                  \nonumber   \\
    &\equiv&B(0) [1+C a^2|t|]~.
\label{4A30}
\end{eqnarray}

 The values for C that we have obtained in our calculations for systems
with $ S_2=S_1=S$ are almost independent of $S/a$.
The values are $ C\simeq -0.54 $ for MM systems
(here is included the case of diquark picture for baryons)
 and $ C\simeq -0.42 $ for
BB systems (three-body picture for hadrons).
  \bigskip

{\bf 5. CHOICE OF PARAMETERS AND COMPARISON WITH EXPERIMENT }

The numerical parametrizations of the total cross section and of the slope
parameter $B$, eqs.(\ref{4A25}) and (\ref{4A26}), are  very convenient
for comparison of our model with
experiment. We first concentrate on elastic $pp$ and $p\bar p$ scattering. In
these channels data are  available over a wide energy
range.
 Since phenomenologically the
Regge-pole parametrization works very well \cite{RRB1,RRB2}, the vaccuum
exchange contribution, to which our models refers, can be extracted as
the pomeron contribution in
a Regge-pole analysis. Donnachie and Landshoff \cite{RRB2} found
that the parametrization
\begin{equation}
\sigma^T_{pom}(pp,\bar pp)=21.70~{\rm mb}~s^{0.0808}
\label{5A1}
\end{equation}
(with s in GeV) works very well over the whole range of
data from $\sqrt s=5$ to
1800~GeV,  so that  we can use this expression for the pomeron contribution.
 According to our convention we choose as energy $\sqrt{s} = 20\ $GeV,
but we wish to enphasize already here that the value of the QCD
parameters, namely the gluon condensate $\langle g^2 FF \rangle$ and the
correlation length $a$ entering the expressions
(\ref{2A13}) and (\ref{3A13}) for the fundamental
correlator are practically independent of the choice of the energy value.
We return to this point at the end of this section.

For the value of the logarithmic slope of the elastic cross-section at
$t=0$ for
$\sqrt s=20\ $GeV we use \cite{RRA3}
\begin{equation}
B=12.47\pm 0.10 \ {\rm GeV}^{-2}~.
\label{5A2}
\end{equation}
This value extrapolates very well, through a Regge amplitude with
  $\alpha'(0)=0.25$~GeV$^{-2}$, to the observed \cite{RRA1}
value  $B=17\ {\rm GeV}^{-2}$ at
$\sqrt s=1800\ $GeV, so that it can be taken confidently as
representative  for the vacuum exchange contribution.

Once the form of the correlator is fixed, we have two parameters in the
model, which are fundamentally related to QCD, namely the gluon
condensate $\langle g^2 FF\rangle$ and the correlation length $a$, and
the extension parameter  $S_H$, determined by phenomenological hadronic
physics. Of course none of these parameters is completely free,
as they appear in other phenomena besides soft high-energy scattering,
and they may be obtained also through lattice calculations. In this section
we take advantage of this independent information to check the consistency
and fix the values of our physical parameters.

   The gluon condensate $<g^2 FF>$ was first determined in 1979 by
Shifman, Vainshtein and Zakharov \cite{RRD4}
   in the framework at QCD sum rules
applied to the charmonium system, with a value  $\langle g^2 FF\rangle =
0.47~{\rm GeV}^4 $. The analysis has been repeated and extended many times,
several authors finding  a range of values extending to considerably larger
values (see Appendix 3). So we here accept as  conservative estimate
the range
\begin{equation}
\langle g^2 FF\rangle=(0.5 \ldots 1.5) \ {\rm GeV}^4 \ .
\label{5A3}
\end{equation}

There are also theoretical uncertainties. Low energy theorems \cite{RRH1}
indicate that, in a world without light quarks, the condensate value
would be larger by
 a factor 2 to 3 than the empirical value of our world with three light
flavours. Therefore in our calculation we must refer to the value
of the gluon condensate which is
valid for a world without light quarks. Our model does not
include dynamical effects of light quarks,
which are treated as external sources moving on given (light-like)
paths. A consequence of the absence of dynamical fermions is that the
fermion (or Wilson loop) renormalization constant is equal to 1. Taking
into account
dynamical fermions not only would change the value of the gluon
condensate but also would lead to a loop renormalization
constant (see eq.(\ref{3A6})),
\begin{equation}
 Z_{\psi}=\bigg<\frac{1}{N_c} \mbox{tr}~W(0,\vec R)\bigg>_A~,
\label{5A4}
\end{equation}
that is smaller than one. Thus in our case we should use a value for the
gluon condensate from a pure gauge theory
\begin{equation}
\big< g^2 FF\big>=(1\ldots 3)\ {\rm GeV}^4 \ .
\label{5A5}
\end{equation}

As already mentioned in sec.~2,  the fundamental correlator $$\langle
F^C_{\mu\nu} (x,0)\ F^D_{\rho\sigma}(0,0)\rangle_A$$ has been
calculated in a pure $SU(3)$  lattice gauge theory by Di Giacomo and
Panagopoulos \cite{RRF1}  using the
cooling method. By this method the high frequency contributions are
frozen out and therefore this correlator is just the one which can be
compared to that of our investigations. The cooling method works very
well for Euclidean distances above about  0.4 fm and
the results obtained in the lattice calculations  show
that the {\it confining}  tensor structure $\{
\delta_{\mu\rho}\delta_{\nu\sigma}-\delta_{\mu\sigma}
\delta_{\nu\rho}\}$ in eq.(\ref{2A13}) is definitely present and even
 dominant, with $\kappa
\sim 3/4$. In the physical range from $r=0.5$ to $r= 0.8$ fm,  the scalar
function $\kappa\langle g^2 FF\rangle D(-r^2/a^2)$ is given by the function

\begin{equation}
\kappa\langle g^2 FF\rangle D(-r^2/a^2)= 24~A \exp(-r/\lambda)~,
\label{5A6}
\end{equation}
with
\begin{equation}
A= 3.6\,~ 10^8~ \Lambda_L^4~~,~~ \lambda = 1/(183 \Lambda_L)~
\mbox{ and }~ \Lambda_L = (0.005 \pm 0.0015)~ \mbox{GeV}~.
\label{5A7}
\end{equation}

Within the given accuracy, the scalar function $D_1$ is proportional to
$D$, and $\kappa \sim 3/4$.
Our choice of the correlator function $D(z^2/a^2)$ given by eq.(\ref{3A21})
 with $n=4$, in the Euclidean region of the lattice results,
namely  for values $1\leq -z^2/a^2
\leq 9 $, is well approximated by an exponential function. We can therefore
determine the gluon condensate and the correlation length $a$
by fitting our expression for the correlator to the one obtained in the
lattice calculation. The values obtained for $\kappa <g^2FF>$ and $a$
depend strongly on the value of the lattice-QCD parameter $\Lambda_L$.
In fig.~5.1 we show our correlator (solid line) together with the
result found in the lattice calculation
in the region $0.5 \leq r \leq 0.8 $ fm, for the choice
$\Lambda_L = 0.0044~\mbox{GeV}$. This fitting leads to the values
$\kappa<g^2 FF> = 1.774 $ GeV$^4$ and $a =0.350 $ fm.


With this value for $a$, our correlator passes through zero for
 $r\approx 1.4$ fm. This change of sign is certainly an artefact of our
special ansatz, but is has no practical consequences,  since
it occurs in a region where the exponential damping makes its
contribution irrelevant anyhow.

As mentioned in sec. 2.2, the evaluation of the (Euclidean) Wilson
loop in the model of the stochastic vacuum yields a relation involving
the condensate
$\kappa \langle g^2 FF\rangle$, the correlation length $a$ and the
string tension $\rho$ (see eq.(\ref{2A18})). For our family of correlators the
integral can be performed analytically (see eq.(\ref{3A24})) and we
obtain  for the case $n=4$
\begin{equation}
\kappa\langle g^2FF\rangle=\frac{81 \pi}{8 a^2} \rho ~.
\label{5A8}
\end{equation}

Finally, we remark that  the extension parameter $S_p$ is not
completely arbitrary,  but
should be in a range of values $S_p=0.5~... 1$ fm, i.e. around
the proton electromagnetic radius.

We now use all this information to analyse our results .

%

In fig.~5.2 we display the relation between the gluon
condensate $<g^2 FF>$ and the
correlation length $a$ obtained from different sources. Fig. 5.2a refers to
the diquark picture and fig. 5.2b to the three-body picture for the proton.
The solid lines show the possible choices of $ <g^2FF>$ and $a$ as
obtained from our model, using eqs.(\ref{4A21}),(\ref{4A23}),
(\ref{4A25}) and (\ref{4A26}) and the
experimental inputs $\sigma^T_{pom} = 34.9 $ mb and $B = 12.47$
GeV$^{-2}$ . To indicate the ranges of values of the proton radius which are
represented in the plots, we mark on these curves the points where
the values of $S_p$ are 0.8 fm (3-body case), 0.9 fm (diquark
case) and 0.6 fm (both cases).

 The dashed lines represents the results
of the lattice calculation \cite{RRF1}, where the largest error
comes from the uncertainty in the lattice QCD parameter $\Lambda_L = 5
\pm 1.5 $ MeV. The values corresponding to some chosen values
of $\Lambda_L$ are marked on this curve. The points for this curve  have
been obtained by fitting the lattice results
to our form for the correlator, as exemplified for a given value
of $\Lambda_L$ in fig.~5.1.

 The dotted lines
represent the relation between the gluon condensate, the correlation
length and the string tension as obtained in the model of the
stochastic vacuum \cite{RRC1,RRC2}; for our form of correlator,
this relation is given by eq.(\ref{5A8}).
The upper and lower curves  correspond respectively to
string tension values $\rho = 0.18 $ GeV$^2$ and $ 0.16$ GeV$^2$.

 As
can be seen from the figures, the constraints from these three
independent sources of information are simultaneously
satisfied in a narrow region, providing a very consistent
picture of soft high-energy $pp$ and $\bar p p$ scattering, for the
following sets of parameters.

1) In the diquark picture
\begin{eqnarray}
&&a=0.350 \mbox{ fm };\quad <g^2FF> = 2.39 \mbox{ GeV}^4\;;\quad S_p = 0.835
\mbox{ fm }; \nonumber \\
&&\Lambda_L = 4.4 \mbox{ MeV }; \quad \rho=0.18~ \mbox{GeV}^2\;.
\label{5A9}
\end{eqnarray}

2) In the three-body picture
\begin{eqnarray}
&&a=0.361 \mbox{ fm };\quad <g^2FF> = 2.08 \mbox{ GeV}^4\;;\quad S_p = 0.730
\mbox{ fm }; \nonumber \\
&&\Lambda_L = 4.2 \mbox{ MeV }; \quad \rho=0.16~ \mbox{GeV}^2\;.
\label{5A10}
\end{eqnarray}

 The calculations described above lead to a determination of
$\kappa <g^2 FF>$ . To obtain the value of the full condensate
quoted above, we
have used $\kappa =0.74$, as determined by the lattice calculation \cite{RRF1}.

The (pure gauge) gluon condensate is well compatible with the canonical
value (see \ref{5A5}). The resulting proton size parameter $S_p$ comes out
quite close to the electromagnetic radius \cite{RRI1} value, which
is $R_p=0.862\pm 0.012$ fm. The lattice parameter $\Lambda_L$ and the
string tension $\rho$ are also in their acceptable ranges.

A very specific feature of the model of the
stochastic vacuum is the dependence of the total cross-section on the
hadron size, even if the latter is large as compared to the correlation
length. The size dependence can
best be tested by comparing the cross-sections for different hadronic
systems. The Donnachie-Landshoff parametrization \cite{RRB2} gives for
the pomeron parts  of $p\pi$ and
$pK$ cross-sections the ratios
\begin{equation}
\sigma_{p\pi}/\sigma_{pp}=0.63 ~\ ; ~\ \sigma_{pK}/\sigma_{p\pi}=
0.87 ~.
\label{5A10A}
\end{equation}
In our treatment the theoretically predicted cross-section ratios
depend on the hadron sizes. We take for the ratios of these
sizes the ratios of the respective electromagnetic radii. The known
values \cite{RRI1}  for these radii are
\begin{equation}
R_p=0.862\pm 0.012~ fm ~, ~R_\pi=0.66\pm 0.01~
fm ~, ~R_K=0.58\pm 0.04~ fm .
\label{5A10B}
\end{equation}
With this input we obtain for
the predicted ratios the results of table~5.1~.

   \vspace{1.5cm}
Table 5.1~Ratios of the pomeron exchange contributions to the total
cross-sections for different processes.

\begin{center}
\begin{tabular}{|c|c|c|c|}  \hline
Cross-section & $p$ picture & $p$ picture & Experimental\\
ratios & diquark & 3-body & values \\ \hline
$\sigma_{p\pi}/\sigma_{pp}$ & $0.66\pm 0.02$ & $0.50 \pm 0.02$ & 0.63\\
$\sigma_{p K }/\sigma_{p\pi }$ & $0.82\pm 0.08$ & $0.82 \pm 0.08$ & 0.87\\
\hline
\end{tabular}
\end{center}
     \vspace{1.0cm}

We see that in the diquark-picture both ratios agree very well with
experiment. In the three-body picture the $\sigma_{p\pi}/\sigma_{pp}$
 ratio comes out too
small. Presumably the assumption that the ratio of hadron radii is
the same as the ratio of the electromagnetic radii  is  an
oversimplification if the assumed structures for the mesons and baryons
are so different,  as when we relate  $\sigma_{p\pi}$ with
$\sigma_{pp}$ with protons in the
three-body picture. This is confirmed by the fact that both pictures
reproduce perfectly well the $\sigma_{p\pi}/\sigma_{pK}$ ratio, where the
effect of the three-body baryon structure is the same in both terms of
the ratio, cancelling out.

With the criterium for the ratios of hadronic radii now fixed,  we can also
calculate the $\pi\pi$ cross-section and we find that for both
the diquark and the three-body picture
the factorization relation
\begin{equation}
\sigma_{\pi\pi}=(\sigma_{p\pi})^2/\sigma_{pp}
\label{5A11}
\end{equation}
is numerically
perfectly fulfilled, as can be verified through the parameter values
given in table 4.2. From the data on the $pp$ and $p\pi$ systems  we thus
predict the pomeron part of the $\pi\pi$ scattering cross-section to be
\begin{equation}
\sigma_{\pi\pi}=8.6~{\rm{mb}}~s^{0.0808}~.
\label{5A12}
\end{equation}
The only experimental result, known to us, on the $\pi\pi$ cross-section
\cite{RRI2}
gives $\sigma_{\pi\pi}=10$~mb at $\sqrt{s} = 4$~GeV, while the above expression
predicts  $9.6$~mb for the pomeron part at this energy.

There are some data available on hyperon-proton scattering \cite{RRA4},
 which  also
fit nicely in the general picture. If we dare to extract a vacuum
exchange part from the few
existing data for $\Sigma p$ scattering, we obtain
\begin{equation}
 \sigma_{p\Sigma^{-}}=19.6~{\rm{mb}}~s^{0.0808}~.
\label{5A13}
\end{equation}
{}From this cross-section we obtain for the ratio of the extension
parameters, in both the diquark and three-body pictures, the value
\begin{equation}
S_{\Sigma^-}/S_p=0.94~,
 \label{5A14}
\end{equation}
which  certainly is a reasonable result.

The size dependence of the slope parameter at $t=0$
is given in parametrized form  by
eq.(\ref{4A26}). In table 5.2 we show the results
of our model for the differences of the slope parameters for
$ p p\,,\Sigma p\,, p\pi \, \mbox{ and } p K $ scattering. The
extension parameters $S_H$ for the different hadrons are assumed
to be proportional to the
electromagnetic radii given in eq.(\ref{5A10B}).
The experimental
numbers for the vacuum exchange part of the slopes were taken from
the analysis by Burq \cite{RRA3} (see their table 7, extrapolated to $t=0$
according
to table 8).  The extrapolation to $t=0$ has little effect on the value
of the slope, for our purposes at this point.

  \vspace{1.5cm}

  Table 5.2~ Differences of the logarithmic slopes of the pomeron part
   at $t=0$ for different processes, in GeV$^{-2}$.

\begin{center}
\begin{tabular}{|c|c|c|c|} \hline
Slopes & Diquark  & 3-Body                 & Experimental \\
        & picture & picture                & values \\ \hline
$B_{pp} -B_{\Sigma p}$ & 0.40 & 0.45 & $ -$   \\
$B_{pp} -B_{\pi p} $ & 1.30   & 1.45 & 2.48 \\
$B_{\pi p}-B_{K p}  $ &  0.43  & 0.34 & 0.34 \\ \hline
\end{tabular}
\end{center}
       \vspace{1.0cm}

   Comparison with the data shows that
the difference of the slopes for $\pi p$ and $ p p $ scattering is
underestimated in our model, while the difference for $\pi p$ and $K p$
is well compatible with experiment.

The parametrization (\ref{4A26}) of the slope parameter
\begin{equation}
  B=1.858\ a^2+0.183\ (S^2_1 + S^2_2)~,
\label{5A15A}
\end{equation}
where $S_1$ and $S_2$ represent the hadron sizes, and $a$
is the correlation length, is a good approximation, within a few percent,
to the results of our model. This parametrization can be compared to that
of a modified Chou-Yang picture \cite{RRJ1}, which
allows also for a quark form factor
\begin{equation}
B={1\over 3} (R^2_q+R^2_1+R^2_2)~,
\label{5A15}
\end{equation}
where $R_q$ and $R_i$ are the electromagnetic radii of the quark and
hadrons respectively. Since in our results $1.858 a^2$ is  about 6 GeV$^{-2}$,
we see that in our model the correlation length $a$
gives a much more important contribution to the slope than the
quark-form factor of the modified Chou-Yang model. These different
predictions should be tested experimentally.

It is a curious result of our treatment that, given the ratios of
the electromagnetic radii of the hadrons, the ratios of the differences
of slope parameters are practically those of simple integers. Thus in
the diquark picture
\begin{equation}
B_{p\pi}-B_{pK}:B_{pp}-B_{p\pi}:B_{pp}-B_{pK}:
            B_{pp}-B_{p\Sigma}= 1:3:4:\frac{9}{10}~.
\label{5A16}
\end{equation}
We remark that the last figure in this sequence of ratios is a bit more
uncertain than the three first ones, because it was determined using the
poor $\Sigma p$ scattering data, and the predicted, not experimentally
measured, ratio $(S_{\Sigma}/S_p)=0.94$.

Our model also predicts a $t$-dependence for the logarithmic slope parameter,
as shown in eq.(\ref{4A30}). In the analysis by Burq \cite{RRA3},
the t-depencence
of the elastic cross-sections has been written in the form
\begin{equation}
\frac{d \sigma}{d t} =\frac{d\sigma}{dt}\bigg\vert_{t=0} e^{[B(0)+ct+dt^2]t}~,
\label{5A17}
\end{equation}
with the parametrization for $B(t)$
\begin{equation}
B(t)= B(0) - 2 c |t| + 3 d |t|^2~.
\label{5A18}
\end{equation}
{}From eq.(\ref{4A30}) and our results for the slope of the slope $C$, we
obtain for this parameter $c$ the values
\begin{eqnarray}
c& = & 10.6 \mbox{ GeV}^{-4} \quad \mbox{in the diquark picture}~,\\
c& = & ~ 8.8 \mbox{ GeV}^{-4} \quad \mbox{in the three-body picture}~.
\label{5A19}
\end{eqnarray}
These values are to be compared with the experimental value $6.8
\pm 0.5 $ GeV$^{-4}$ obtained by Burq \cite{RRA3}. However we should not
overestimate the significance of this comparison. On the theoretical
side this slope of the slope depends strongly on the precise form
of the profile function $J(b)$, while the experimental results are not
accurate and  may still be contaminated by Coulomb interference.

If we consider hadron-hadron scattering for two hadrons of equal sizes, we
 can eliminate the hadron radius between the parametrized forms for
$\sigma^T$ and $B$, and obtain the relation
\begin{equation}
\sigma^T_{pom} = \alpha \gamma^{-\beta/\delta} (\kappa <g^2 FF>)^2
a^{10-2\beta/\delta} (B - 1.858 a^2)^{\beta/\delta}~,
\label{5A20}
\end{equation}
 where the values for
the model dependent parameters $\alpha, \beta, \gamma, \delta $ are given in
table 4.2 .

If we fix  $\kappa \langle g^2~FF \rangle$ and $a$ at one given energy
(e.g. $\sqrt{s}= 20$~GeV), we obtain through eq.(\ref{5A20}) a parameter-free
relation between the total cross-section $\sigma^T_{pom}$ and the slope
parameter $B$.

In fig.~5.3  we display $\sigma^T_{pom}$ against $B$
as given by  eq.(\ref{5A20}), for both diquark  and
three-body pictures for the proton,  using the sets of parameters
given in table 4.2 .
In the same figure we also show the relation
\begin{equation}
  \sigma^T_{Regge} = \sigma^T_0~ e^{{0.1616} (B-B_0)}~,
\label{5A20A}
\end{equation}
obtained from a Regge amplitude
using the slope of the pomeron trajectory
$\alpha'(0)_{pom}=0.25~$GeV$^{-2}$,
and as input at $\sqrt{s}=20 {\hbox{\rm GeV}}$ the values
$\sigma^T_0=35$~mb, and $B_0=12.47~{\hbox {\rm GeV}}^2~.$
The experimental data \cite{RRA1} at 540 GeV and 1800 GeV  are marked
in the plot, together with the input data at 20 GeV. We
observe that our relation (\ref{5A20}),  which contains no free
parameters, describes the experimentally observed relation between $B$
and $\sigma^T$ astonishingly well. Besides, there is a surprising
agreement between the Regge parametrization line and our results. It
must be remarked  that the constant
term $1.858~a^2$ in our expression for the slope $B$ is important  for
this good agreement with experiment.

The application of our results to different energies implies a very
slow energy
dependence of the hadronic radii. An explicit relation is obtained if
we bring into eqs.(\ref{4A21}) and (\ref{4A27}) a parametrization for the
energy
dependence of the total cross-sections, such as the Donnachie-Landshoff
\cite {RRB2} form of eq.(\ref{5A1}). We thus obtain for the proton radius
\begin{equation}
S_p(s)= a\bigg[\frac{\kappa\langle g^2 FF \rangle^2 a^{10}(21.7 {\rm mb})}
          {0.00626} \bigg]^{1/3.090}~s^{0.0808/3.090}
\label{5A19A}
\end{equation}
in the diquark picture, and
\begin{equation}
S_p(s)= a\bigg[\frac{\kappa\langle g^2 FF \rangle^2 a^{10}(21.7 {\rm mb})}
          {0.00881} \bigg]^{1/3.277}~s^{0.0808/3.277}
\label{5A19B}
\end{equation}
in the 3-body configuration case. The values
thus obtained for $S_p$ are in the region of the proton electromagnetic
radius.


{\bf  6. CONCLUSIONS }

We have obtained a very consistent description of the data on soft high-
energy scattering. Our basic assumption is that the low frequency, i.e.
nonperturbative, contributions to the scattering amplitudes can be
approximated by a Gaussian process. The parameters determining the observable
quantities are
the gluon condensate and the correlation length of the
vacuum field fluctuations. These parameters occur also in completely
different connections (SVZ sum rules, lattice calculations, low energy
hadron spectroscopy), and all the conditions posed on them can be consistently
satisfied. The third parameter entering our calculations is the
transverse hadron
size, which may be related to the electromagnetic radius.

Gauge invariance is observed on each step of the calculation, which is
based on a formalism for loop-loop scattering, rather than on
a quark-quark scattering picture. In this way even finite distance
correlations of the vacuum field tensor lead to long range correlations
that are the common source of confinement,  and to a dependence of
the total
scattering cross-section on the hadron size. This
mechanism can be interpreted as a string-string interaction in hadronic
scattering.

The size dependence of the cross-sections leads to a
natural explanation for the experimentally observed flavour dependence
of the total cross-section. The ratio
between the pion-proton and (anti)proton-proton cross-sections emerges
in our model as a consequence of the different hadron
sizes, and the factor 2/3 comes out as a consequence of the ratio of the
electromagnetic radii,  and not from quark additivity.

  The size dependence of the slope parameter, i.e. the logarithmic slope
of the differential elastic cross-section, can be parametrized in a form
similar to that of a modified Chou-Yang model \cite{RRJ1} with a finite
quark radius, which in our model appears as a correlation length.

Elimination of the extension parameter yields a parameter-free relation
between the total cross-section and the slope of the elastic cross-section
which agrees very well with experiment.

The investigations described in this paper can be extended in many
directions. In the present calculations only one of the two possible tensor
structures determining the low frequency contributions is taken into
account. The inclusion of the second term, which could also describe
perturbative effects (and even Coulomb interaction), would  pose no
important technical problems. Furthermore, we have restricted  ourselves
to the lowest order
non-vanishing contribution, which is quadratic in the gluonic
correlator. We have checked that this is justified for the total
cross-section and the slope parameter. But our amplitude is purely
imaginary, and quantities like the
$\rho$-parameter (the ratio of the real to the imaginary parts of the
elastic scattering amplitude) can only be described if we go one
further order in the contributions to the correlator.

A further important step to be developed is the test of the importance
of the factorization implied by the assumption of a Gaussian process.
This is certainly crucial in the present investigation, with possible
consequences for the phenomenological analysis, but it remains to be seen
which of the more general features depend on this approximation.


It would of course be highly desirable improve the present model in order to
describe the dependence of the cross-section with the energy. However,
 although we cannot obtain the (slow) rise of the
total cross-section with the energy, we can eliminate the hadron radius
parameter $S$ and obtain a parameter free relation between the total cross-
section and the slope parameter $B$, as shown in fig. 5.3 .

\bigskip

{\bf Acknowledgements}

We thank Adriano Di Giacomo, Peter Landshoff, Otto Nachtmann and Yuri
Simonov for many fruitful discussions and constructive critical remarks.
We are also indebted to Otto Nachtmann for a critical reading of the
manuscript.

Part of this work has been done when two of the authors were visiting CERN.
They (H.G.D. and E.F.) wish to thank the Theory Division for the hospitality.
The same authors also thank CNPq (Brasil) and DAAD (Federal Republic
of Germany) for the financial support. One of the authors (A.K.)
gratefully acknowledges the support received from  Studienstiftung des
Deutschen Volkes.
\newpage

{\bf  Appendix 1 . THE CORRELATION FUNCTIONS    }

 In eqs.(\ref{2A13}) and (\ref{3A13}) there appear two independent
arbitrary scalar functions, $D(z^2/a^2)$ and $D_1(z^2/a^2)$, which
are supposed to fall off at large distances
with  characteristic lengths $a$, called correlation lengths, and
must have forms that can be analytically continued  from Euclidean
to Minkowski space-time descriptions of field theory. If the
expression for the correlator is used
in Euclidean QCD, the scalar function $D(z^2/a^2)$,
which is zero in Abelian theories (if there are no monopoles)
leads to a linearly rising potential, namely  to confinement. QCD lattice
calculations  have shown  that  the dominant contribution to the correlator
actually comes from the term with $D(z^2/a^2)$ , namely $\kappa\simeq 1 $.
Besides that, in Appendix 2 we  show that the non-confining part $D_1$ of
the correlator has much less influence on the values of the eikonal
functions of high-energy scattering. Consequently, in the
present work we take into account only  the confining term
$D(z^2/a^2)$, with the weight $\kappa$ determined by the lattice calculations,
  neglecting the effect of $ D_1(z^2/a^2)$  altogether.

         We thus concentrate on $D(z^2/a^2)$, and take as an ansatz the
family of functions
\begin{equation}
D^{(n)}(\xi^2)=-6i\int{\frac {d^4k}{(2\pi)^4}~\frac{A_nk^2}{(k^2-1)^n}
              ~\exp{(-ik\xi/\lambda_n)} } ~,~~~ n\geq 4~,
\label{7A1}
\end{equation}
where
\begin{equation}
        \xi = z/a ~,
\label{7A2}
\end{equation}
$a$ is the characteristic correlation length,
and  the constants $A_n$  and $\lambda_n$  are to be fixed by normalization.
It is convenient to absorb  $\lambda_n$ into $k$ through
 $k/{\lambda_n}\rightarrow k$ ; then $\xi$,$k$ and $\lambda_n$ are all
dimensionless. In the Euclidean metric
\begin{equation}
   -i d^{4}k = d^{4}K~,~K_4=i k_0~,~k^2 = - K^2 = -(|\vec K|^2+K_4^2)~,
\label{7A3}
\end{equation}
for space-like vectors
\begin{equation}
\xi(0,\vec{\xi}=\sqrt{-\xi ^2} \vec e_3)~,~ \xi^2=-|\vec{\xi}|^2~,
\label{7A4}
\end{equation}
we have
\begin{equation}
D^{(n)}(\xi^2)=D^{(n)}(-|\vec\xi|^2)
=6(-1)^{n+1}\int{\frac {d^4K}{(2\pi)^4}~\frac{A_n\lambda_n^{6} K^2}
{(\lambda_n^2 K^2+1)^n}
          ~\exp{({i\vec K}\cdot {\vec\xi})} }~.
\label{7A5}
\end{equation}

The constants $A_n$ and $\lambda_n$  can be fixed by the normalization
conditions
\begin{equation}
     D^{(n)}(0)=1 ~~ , ~~
       \int_{0}^{\infty}d(|\vec{\xi}|)~D^{(n)}(-|\vec{\xi}|^2) = 1 ~.
\label{7A6}
\end{equation}
The second of these two relations has the role of a definition for
the correlation length.

These calculations can be made analytically, leading to
\begin{equation}
       A_n = (-1)^{n+1} \frac{4\pi^2}{3}(n-1)(n-2)(n-3)~ ,
\label{7A7}
\end{equation}
and
\begin{equation}
\lambda_n = \frac{1}{\rho_n} =
 \frac {4}{3\sqrt \pi}\frac{\Gamma(n-3)}{\Gamma(n-5/2)}~.
\label{7A8}
\end{equation}

  For simplicity of notation, from now on we use more often
$\rho_n=1/\lambda_n$, instead of $\lambda_n$.

  The integrations in eq.(\ref{7A5}) can be performed analytically. We
obtain
\begin{equation}
D^{(n)}(-|\vec\xi|^2)=\frac{(-1)^{n+1}3 A_n}{\pi^2 2^{n-1}\Gamma(n)}
 (\rho_n|\vec\xi|)^{n-3}
 \bigg[(n-1)K_{n-3}(\rho_n|\vec\xi|)
      -\frac{1}{2}(\rho_n|\vec\xi|)
                 K_{n-2}(\rho_n|\vec\xi|)\bigg]~,
\label{7A9}
\end{equation}
 which is the general form for the class of functions considered.
 $K_\nu (x)$ is the modified Bessel function.

    Alternative useful forms to eq.(\ref{7A9}) are
\begin{equation}
D^{(n)}(-|\vec\xi|^2)=\frac{(-1)^{n+1}3 A_n}{\pi^2 2^{n}\Gamma(n)}
  \bigg(\rho_n|\vec\xi|\bigg)^{-3}
 \frac{d}{d(\rho_n|\vec\xi|)}\bigg[(\rho_n|\vec\xi|)^{n+1}
 K_{n-3}(\rho_n|\vec\xi|)\bigg]~,
\label{7A10}
\end{equation}
and
\begin{eqnarray}
D^{(n)}(-|\vec\xi|^2)
 =\frac{(-1)^{n}3 A_n}{\pi^2 2^{n}\Gamma(n)}
  (\rho_n|\vec\xi|)^{-3}
 \frac{d}{d(\rho_n|\vec\xi|)} \bigg[(\rho_n|\vec\xi|)^{3}
  \frac{d}{d(\rho_n|\vec\xi|)} [
                   (\rho_n|\vec\xi|)^{n-2}
 K_{n-2}(\rho_n|\vec\xi|) ] \bigg]~.
\label{7A11}
\end{eqnarray}

        These correlation functions are negative at large
distances, behaving like
\begin{equation}
D^{(n)}(-|\vec\xi|^2)\simeq -\frac {\sqrt{\pi/2}}
        {2^{n-2}\Gamma(n-3)}(\rho_n|\vec\xi|)^{n-5/2}
       \exp{(-\rho_n|\vec\xi|)}~.
\label{7A12}
\end{equation}

\bigskip
       After the limits are taken, which make the long sides of the
rectangular Wilson loops tend to $\pm \infty$ in the $x_3$ direction, the
remaining variables in the
integrands  are coordinates of points in the transverse plane.
The distances $z$ between such points enter in the final expressions
for the eikonal functions $\chi$
as arguments of the two-dimensional inverse Fourier transform
of $ \tilde D(k^2) $ in eq.(\ref{3A21}), which  is defined by

\begin{equation}
                   {\cal F}_{2}^{(n)}(-|\vec\xi|^2)
    = - \frac{1}{(2\pi)^2} \int{  {d^2\vec K_\bot}~
         \frac{6A_n\lambda_n^{6}|\vec K_\bot |^2 }
         { (-\lambda_n^2|\vec K_\bot|^2-1)^n}
          ~\exp{({i\vec K_\bot}\cdot {\vec\xi})} }~.
\label{7A13}
\end{equation}
     where $\vec \xi$
      is any two-dimensional vector of the transverse plane.

      Using the values for $A_n$ and $\rho_n$ written  above, we
obtain
\begin{equation}
              {\cal F}_{2}^{(n)}(-|\vec\xi|^2)
 =  \frac {2^{(8-n)}\Gamma(n-3)}{9[\Gamma(n-5/2)]^2}
 (\rho_n|\vec\xi|)^{n-2}\bigg[(n-1) K_{n-2}(\rho_n|\vec\xi|)
-{1\over 2}(\rho_n|\vec \xi|) K_{n-1}(\rho_n|\vec\xi|) \bigg]~.
\label{7A14}
\end{equation}
     Important alternative forms are
\begin{equation}
              {\cal F}_{2}^{(n)}(-|\vec\xi|^2)
 =  \frac {2^{(7-n)}\Gamma(n-3)}{9 [\Gamma(n-5/2)]^2}
 (\rho_n|\vec\xi|)^{-1}\frac{d}{d(\rho_n|\vec\xi|)}\bigg[
   (\rho_n|\vec\xi|)^{n} K_{n-2}(\rho_n|\vec\xi|) \bigg]~,
\label{7A15}
\end{equation}
and
\begin{equation}
              {\cal F}_{2}^{(n)}(-|\vec\xi|^2)
 = -\frac{2^{(7-n)}\Gamma(n-3)}{9[\Gamma(n-5/2)]^2}
 (\rho_n|\vec\xi|)^{-1}\frac{d}{d(\rho_n|\vec\xi|)}\bigg[
   (\rho_n|\vec\xi|)\frac{d}{d(\rho_n|\vec\xi|)}\bigg(
 (\rho_n|\vec\xi|)^{n-1}
   K_{n-1}(\rho_n|\vec\xi|) \bigg)\bigg]~,
\label{7A16}
\end{equation}
 or
\begin{equation}
              {\cal F}_{2}^{(n)}(-|\vec\xi|^2)
 = - \frac {2^{(7-n)}\Gamma(n-3)}{9 (\Gamma(n-5/2))^2}
\Delta_2\psi^{(n)}(\rho_n|\vec\xi|)~,
\label{7A17}
\end{equation}
where
\begin{equation}
\psi^{(n)}(\rho_n|\vec\xi|)\equiv
             (\rho_n|\vec\xi|)^{n-1} K_{n-1}(\rho_n|\vec\xi|)~,
\label{7A18}
\end{equation}
and $\Delta_2$ is the 2-dimensional Laplacian operator.
This Laplacian form
is important in our calculation, as it allows lowering the order
of the integrations, through Gauss theorem.

  The moments of the
 $D^{(n)}(-|\vec\xi|^2)$ and  ${\cal F}_{2}^{(n)}(-|\vec\xi|^2)$ functions
can be readily obtained from eqs.(\ref{7A10}) and (\ref{7A15})
respectively. For $D^{(n)}$ we obtain
\begin{eqnarray}
{\cal M}_p[D^{n}(-|\vec \xi|^2)]= \int_0^\infty  (\rho_n |\vec\xi|)^p
    ~ D^{(n)}(-|\vec \xi|^2)   d(\rho_n |\vec\xi|)  \nonumber \\
   = 2^{p-2} (3-p)\frac{\rho_n^{-p+1}}{ \Gamma(n-3)}
    \Gamma\bigg(\frac{p+1}{2}\bigg)\Gamma\bigg(n+\frac{p-5}{2}\bigg)
\label{7A19}
\end{eqnarray}
where we observe that $ {\cal M}_3[D^{n}(-|\vec \xi|^2)]=0 $
and that all moments higher than $p=3$ are negative, for every $n$.
Thus all correlation functions have a zero.

   For  ${\cal F}_{2}^{(n)}$  we have the moments
\begin{eqnarray}
{\cal M}_p[{\cal F}_{2}^{(n)}(-|\vec\xi|^2)]=
 \int_0^\infty  (\rho_n |\vec\xi|)^p
  ~ {\cal F}_{2}^{(n)}(-|\vec\xi|^2) d(\rho_n |\vec\xi|)  \nonumber \\
   ={2^{(p+4)}\over 9}(1-p)\frac{\rho_n^{-p+1}\Gamma(n-3)}{[\Gamma(n)]^2}
    \Gamma\bigg(\frac{p+1}{2}\bigg)\Gamma\bigg(n+\frac{p-3}{2}\bigg)~.
\label{7A20}
\end{eqnarray}
Thus for every $n$ the moment $p=1$ vanishes, and all higher moments are
negative. Thus all  ${\cal F}_{2}^{(n)}$ functions pass through  zero.

  We must choose the value of n in order to fix a specific form
 of the  ansatz for the correlation function. Actually, the dependence
of the final results on this particular choice has been tested \cite{RRG2},
and found to be not very marked. The reason is that all correlation
functions are normalized to 1 at the origin, and decrease
exponentially at large distances. It is enough that the chosen function
falls monotonicaly and smoothly
in the range of physical influence (up to about one fermi, say), and
there cannot be much difference
in the results obtained using different analytical forms. Of course
there will be differences in the specific values given to the correlation
length parameter $a$, due to the different values of the multiplicative
factor $\rho_n$ in the argument of the exponential behaviour
(see eq.(\ref{7A12})), but such
differences can be taken into account and absorbed when different forms
of correlation functions are compared.

   In the present work we make the choice that n=4, which in the Euclidean
region leads to a good representation of the lattice calculations
\cite{RRF1}. We then have for the correlation function
\begin{equation}
D^{(4)}(-|\vec\xi|^2)=  (\rho_4|\vec\xi|)
 \bigg[K_{1}(\rho_4|\vec\xi|)
      -\frac{1}{4}(\rho_4 |\vec\xi|) K_{0}(\rho_4 |\vec \xi|)\bigg]
\label{7A21}
\end{equation}
and for the 2-dimensional inverse Fourier transform
\begin{eqnarray}
              {\cal F}_{2}^{(4)}(-|\vec\xi|^2)
 =\frac{32}{9\pi}(\rho_4|\vec\xi|)^2 \bigg[ 2 K_{0}(\rho_4|\vec\xi|)
-\bigg(\frac{4}{\rho_4|\vec \xi|}-
\rho_4|\vec \xi|\bigg) K_1(\rho_4|\vec\xi|) \bigg]
       \nonumber  \\
              = -\frac{32}{9\pi}\Delta_2
            \big[(\rho_4|\vec\xi|)^{3} K_{3}(\rho_4|\vec\xi|)\big]~,
\label{7A22}
\end{eqnarray}
where
\begin{equation}
              \rho_4= \frac{3\pi}{8}~.
\label{7A23}
\end{equation}

     The functions $D^{(4)}(-|\vec\xi|^2)$, ${\cal F}_{2}^{(4)}(-|\vec\xi|^2)$
and  $\psi^{(4)}(\rho_4|\vec\xi|)$  are represented in figs. (7.1),(7.2)
and (7.3), against
the variable $x=\rho_4|\vec\xi|$. The correlation function
$D^{(4)}(-|\vec\xi|^2)$ has a zero at $\rho_4 |\vec\xi|=4.43 $
while   ${\cal F}_{2}^{(4)}(-|\vec\xi|^2)$ has a zero at
 $\rho_4 |\vec\xi|=3.05$. As we will see from our final results,
the locations of these zeros are beyond the range of  physical
influence.


      Practical representations for these functions, that are
important for the numerical work, can be obtained. As a tool to obtain
the parameters in approximate representations, we may use the moments of the
functions, which are explicitly given by eqs.(\ref{7A19}) and (\ref{7A20}).
The function
\begin{equation}
\psi^{(4)}(x)=x^3 K_3(x)~,
\label{7A23A}
\end{equation}
where $x=\rho_4|\vec \xi|$~, enters
in heavy numerical computations to produce the eikonal
functions. It appears in eq.(\ref{4A7}) as
 $$ \frac{d\psi}{dx}=-x^3 K_2(x)~. $$
Let us write
\begin{equation}
     \psi^{(4)}(x)\simeq 8~[1+x+a_1 x^2+a_2 x^3+a_3 x_4]~e^{-x}~,
\label{7A24}
\end{equation}
that satisfies the constraints at the origin $\psi(0)=8$ and
$\psi^{\prime}(0)=0$~. We then have
\begin{eqnarray}
 {\cal F}_{2}^{(4)}(-|\vec\xi|^2)
     \simeq -\frac{2^8}{9\pi}\bigg[(-2+4a_1)+(1-5a_1+9a_2)x  \nonumber \\
+(a_1-7a_2+16a_3)x^2+(a_2-9a_3)x^3+a_3 x^4\bigg]~e^{-x}~.
\label{7A25}
\end{eqnarray}

Using  properties of $ {\cal F}_{2}^{(4)}(-|\vec\xi|^2)$, we obtain
\begin{eqnarray}
      {\cal F}_{2}^{(4)}(x=0)=2^7/(9\pi)~ & \Longrightarrow &~
                    a_1=3/8 ~,    \nonumber \\
\bigg[ \frac{d}{dx}{\cal F}_{2}^{(4)}\bigg](x=0)=0~ & \Longrightarrow &~
            a_2=1/24~,       \nonumber \\
 \int_0^\infty {\cal F}_{2}^{(4)}dx=16/3 ~ & \Longrightarrow &~
                                a_3={1\over8}({7\over3}-{3\pi\over4}) ~,
      \nonumber \\
 \int_0^\infty x~{\cal F}_{2}^{(4)}dx=0~ & \Longrightarrow &~
                               ~\hbox{\rm identity}~.
\label{7A26}
\end{eqnarray}
There are no more free parameters. The next moment
$$ \int_0^\infty x^2~{\cal F}_{2}^{(4)}dx=-80/3 $$
is reproduced in good approximation.

    The representations (\ref{7A24}) and (\ref{7A25}) are excellent for
our purposes, and can be safely used, reducing
substantially the computation time.

   {\bf Appendix 2.~ Influence of the confining and non-confining correlators
 on high-energy scattering }

We have seen in subsec. 2.2 that only one of the tensor structures
present in eq.(\ref{3A13}) leads to confinement, namely the part of the
correlator whose scalar function is denoted by D. If $\kappa =1$ this is
the only existing contribution, and since it has the property of
leading to confinement, we refer to it as the
{\em confining case }. The other
tensor structure, which for $\kappa$ = 0 is the only one contributing,  is
referred to as the {\em non-confining case}.
  Comparing the effects of the two functions, we put
\begin{equation}
-{1\over 2}k^2  {d\over dk^2}\widetilde D_1(k^2) = \widetilde D(k^2)~.
\label{3A25}
\end{equation}
in order to have the same spatial behaviour
for their contributions to the correlator.


In fig. 8.1 we represent the reduced
eikonal function $\widetilde\chi(\vec b, \vec R(1,1), \vec R(2,1))$ for
the case where $\vec b, \vec R(1,1)$, and $\vec R(2,1)$ are all parallel
and with a ratio  $|\vec R|/a = 2$. The eikonal function $\widetilde\chi$
is plotted as a function of $b/a$, i.e. the impact parameter in units of the
correlation length $a$. In the
confining case the eikonal function is approximately proportional to
the overlap region of the two loops in the transverse plane. In the
non-confining case there is a large contribution if both quarks or
antiquarks from loop 1 and loop 2  coincide, i.e. for $b=0$ ; the
contribution is smaller if there is coincidence of only one particle
(quark or antiquark) from of each loop.
Thus in the confining case we have typically a true string-string
interaction, while in the non-confining case we have a quark-quark
interaction.


In fig. 8.2  we
show the strong dependence of the eikonal function $\widetilde\chi$ on the
orientation of the loops for the confining case. The figure shows
$\widetilde\chi$ as a function of $b/a$, for the case that the loops
are parallel
in the transverse plane i.e. $\theta_1 = \theta_2 = \theta$. The
common angle $\theta$ is a parameter, running from $0^o$
to $80^o$ in steps of $20^o$. In order to strengthen and make very  visible
the effect, we have chosen a large value $|\vec R|/a = 10$. From the
figure we observe that the loops
interact as one-dimensional objects in the transverse plane.


 In fig. 8.3
we show the cross-section (in arbitrary units) as a function of the
hadron extension $S/a$, using a Gaussian wave function
$\sqrt{2/\pi} (1/S) \exp(-R^2/S^2)$.
For $S/a \leq 1$, the two cases nearly coincide, but for larger values
the cross-section in the non-confining case stays nearly constant, whereas the
cross-section for the confining case continues to increase
approximately like $(S/a)^3$. Thus we have quark additivity only for the
non-confining case. Since the correlator for the confining case can
only be present in a non-Abelian theory, an Abelian model will always
yield quark additivity if $(S/a) \geq 1$ holds \cite{RRD5}.

\bigskip
\vspace{1.5cm}

      {\bf Appendix 3. ~ The value of the gluon condensate $<g^2 FF>$}.

The most accurate determination of the gluon condensate $<g^2 FF>$
comes from sum rule analyses of the charmonium system. In their
original paper on sum rules, Shifman, Vainshtein and Zakharov \cite{RRD4}
obtained the value
\begin{equation}
\langle g^2 FF \rangle = 0.47~\mbox{GeV}^4~.
\label{9A1}
\end{equation}
This analysis was extended \cite{RRK1}, yielding, taken into
account all observed ground states of given quantum numbers in the
charmonium system the range of values
\begin{equation}
0.51 ~\mbox{GeV}^4~ \leq~\langle g^2 FF \rangle~ \leq 0.79 ~\mbox{GeV}^4~.
\label{9A2}
\end{equation}

Other analyses \cite{RRK2,RRK3,RRK4} yield considerable
higher bounds for the gluon condensate. The main uncertainties are
\cite{RRK5} the error in the pole mass
of the charmed quark, radiative corrections and the contributions of
higher condensates. The last point is the most difficult to control
 \cite{RRK3}. Models
indicate that they could increase considerably the value of the gluon
condensate \cite{RRK4,RRK6}. A conservative estimate for the
gluon condensate is
\begin{equation}
\langle g^2 FF \rangle = (0.95 \pm 0.45)~\mbox{GeV}^4~.
\label{9A3}
\end{equation}
Novikov \cite{RRK7} has however argued convincingly against the larger values
(say larger than 0.8 GeV$^4$) for the reason that the analyses yielding
large upper bounds for the gluon condensate either rely on only one specific
channel, or are based on particular models. A non-diagonal sum rule analysis
of the matrix element $\langle 0|{\bf \widetilde A}_\mu   |\pi\rangle $, with
\begin{equation}
      {\bf \widetilde A}_\mu= \frac{1}{2} g~
\epsilon_{\rho\mu\alpha\beta}~ \bar d~ \gamma_\rho~ {\bf F}_{\alpha\beta}~u ~,
\label{9A4}
\end{equation}
that is particularly sensitive to the gluon condensate, yields also a
low value.

There is however another theoretical difficulty. In a world without
light quarks the value of the gluon condensate could be considerably
larger. A low-energy theorem has been derived \cite{RRH1} relating
the change of the gluon condensate with respect to (light) quark masses
with the quark condensate
\begin{equation}
\frac{d}{dm_q} <g^2 FF> = \frac{96 \pi^2}{\beta_1} <\bar q q>~,
\label{9A5}
\end{equation}
where $<\bar q q>$ is the light quark condensate and $\beta_1$ is the
first coefficient of the Gell-Mann Low function. For SU(3) the factor is
large, $ {96 \pi^2}/{\beta_1}\approx 100$.

Since the quark condensate $<\bar q q>$ is negative, an increase of the
quark mass leads to an increase of the gluon condensate. In a world of
pure gauge fields the gluon condensate would thus have a larger value.
This {\it pure gauge} value is obtained from the {\it empiric} one, which
refers to a world with three light flavours, by taking all masses to
infinity, i.e. by integrating eq.(\ref{9A5}) up to a value of $m_q$ where
the quarks decouple. The authors of the theorem \cite{RRH1} estimate
that the pure gauge value is about two to three times higher than the
empirical value. As argued in sec. 5, in our model we must use the pure
gauge value of the gluon condensate.


\newpage
{\bf Figure Captions}
\bigskip

{ Figure  3.1~
Wilson loops formed by the paths of quarks and antiquarks inside
two mesons. The impact vector $\vec b$ is the distance vector between
the middle lines of the two loops. $\vec R_1$ and $\vec R_2$ are the
vectors in the transverse plane from the middle lines to the quark
lines of meson 1 and
2 respectively. For the antiquarks the corresponding vectors are $-\vec
R_1$ and $-\vec R_2$. The front lines of the loops guarantee that
the mesons behave
as  singlets under local gauge transformations.}

\bigskip

Figure 3.2~ Representation of the non-Abelian Stokes theorem. The contour
integral running from $w$ to $x_{(1)}$ then to $x_{(2)}$ and back to $w$
is deformed in order to become a surface integral. Here  $x_{(1)}$ and
$x_{(2)}$ represent the coordinates of the quark and antiquark in a meson,
respectively.
\bigskip

Figure 3.3 ~Tilted perspective view of the surfaces $S_1$ and $S_2$
obtained from
the line integrals along the Wilson loops $\partial S_1$ and $\partial
S_2$ after appying the non-Abelian Stokes theorem. The line
$ \overline{B_{i1}A_{i1} }$ is the quark path line and
$\overline{B_{i2}A_{i2}}$ is
the antiquark path line of meson $i$. C is the reference point (with
coordinates $w$ in the non-Abelian Stokes theorem).
\bigskip

Figure 3.4 ~Wilson loops (without traces) describing a baryon path in the
three-body picture. The line from $x_{(i)}$ to $x'_{(i)}$ represents the
path for the quark labelled $i$. The line from $x_h$ to $x'_{h}$ is the
path for the central point of the baryon.
 \bigskip

 Figure 4.1~
       Geometrical variables of the transverse plane, which
       enter in the calculation of the eikonal function
       for meson--meson scattering. The points $C_1$ and $C_2$ are the
       meson centres. In the integration, $P_2$ runs
       along the vector $\vec Q(2,1)$, changing
       the length $z$, which is the argument of the correlator
       characteristic function. In analogous terms, points  $P_1$,
       $\bar P_1$ and $\bar P_2$ run along $\vec Q(1,1)$, $\vec Q(1,2)$ and
       $\vec Q(2,2)$. This explains the four terms that appear in
       eq.(\ref {4A6}) .
\bigskip

Figure 4.2 ~Geometrical variables for the calculation of the eikonal
function for meson-baryon scattering. Notation analogous to fig. 4.1~.

\bigskip

 Figure 4.3 ~
     Dimensionless function $J(b/a)$, where $b$ is the impact
parameter, and $a$ is the correlation length. The three
represented cases refer to baryon--baryon ($BB$), meson--baryon ($MB$)
and meson--meson ($MM$) amplitudes, with the same values
($S_1=S_2=S=4a$) for the hadron extension parameters.
 \bigskip

 Figure 4.4 ~
     Function $I_0(S/a)$ representing the dependence of the total
cross-section on the extension parameter S (in units of the correlation
length $a$) of the hadron wave-function. M represents a meson-like
hadronic configuration ( $q\bar q$ for mesons, quark-diquark for
baryons), and B represents a 3-body star-like picture for baryons.
 We have here used $S_1=S_2=S$ in all cases.

\bigskip

 Figure 4.5 ~
    Same as for previous figure, for the quantity $K(S/a)$, which
represents the slope parameter $B$ divided by $a^2$.
\bigskip

 Figure 5.1 ~     Demonstration of the
determination of the QCD parameters through the  fitting of our correlator
to the lattice calculation results \cite{RRF1}, for a given value of
$\Lambda_L$. Dashed line: lattice
results for $\Lambda_L = 4.4 $ MeV. The arrows
indicate the range inside which the lattice calculations were made.
Solid line: our fitted correlator, with the fitted values $a=0.350$~fm and
$\kappa \langle g^2 FF\rangle D(-|\vec \xi|^2)= 1.772~$GeV$^4$ at
$|\vec \xi|= 0$.
\bigskip

 Figure  5.2a~
Constraints on the values of  $\kappa \langle g^2 FF\rangle$ and of the
correlation length $a$. Solid line: Our model of high energy
scattering, with $\sigma^T_{pp} = 35$ mb and $B=12.47$ GeV$^{-2}$,
using a diquark picture for the proton ;
dashed line : fit of our correlator to the lattice  calculation \cite{RRF1};
dotted lines : relation obtained
from string tension $\rho$, eq.(\ref{5A8}), the upper curve
corresponding to $\rho=0.18 $ GeV$^2$, and
the lower one to $\rho=0.16 $ GeV$^2$.

\bigskip

Figure 5.2b~
Same as fig. 5.2a, for the case of a 3-body picture for the proton.

\bigskip

 Figure 5.3 ~
Relation between the total cross section and the slope parameter $B$.
Solid lines: predictions given by eq.(\ref{5A20}), obtained from
our model by eliminating the hadronic size, in the cases of diquark and
3-body pictures for the proton  ;
dashed line: relation obtained from the Regge amplitude with
$\alpha(0) = 1.0808$ and $\alpha'(0) = 0.25~$GeV$^{-2}$; star : input
data at $\sqrt {s}= 19~$GeV ; crosses : experimental data at 541 and
 1800~ GeV.
\bigskip

 Figure 7.1~ Correlation function $ D^{(4)}(-|\vec\xi|^2)$~, given by
eq.(\ref{7A21}), against the variable $ x= \rho_4 |\vec \xi|
= ({3\pi}/{8})~(z/a)~$, where $z$ is the physical distance, and
$a$ is the correlation length. The correlation function is normalized
to one at the origin.

\bigskip

 Figure 7.2~ Two-Dimensional Fourier transform
${\cal F}_{2}^{(4)}(-|\vec\xi|^2)~$,
given by eq.(\ref{7A22}), against $ x= \rho_4 |\vec \xi|
= ({3\pi}/{8})~(z/a)~$.
\bigskip

 Figure  7.3~ Function $\psi^{(4)}(x)=x^3 K_3(x)~$, given by eq.(\ref{7A23}),
and represented approximately through eq.(\ref{7A24}).

\bigskip

Figure 8.1~ Comparison of the values of the reduced eikonal functions
$\widetilde\chi~$ for the extreme pure confining ($\kappa=1$) and
pure non-confining ($\kappa=0$) cases, in a situation where the three
vectors $\vec b$ , $\vec R(1,1)$ , and $\vec R(2,1)$ are parallel.

\bigskip

Figure 8.2~ Angular dependence of the reduced eikonal function
$\widetilde \chi$ for the purely confining case ($\kappa=1$). The figure
shows results for configurations where the two interacting loops are
parallel ($\theta_1=\theta_2=\theta$) and
$|\vec R_1| = |\vec R_2| = 10 a$~. The angle $\theta$ varies from $0^o$
to $80^o$ in steps of $20^o$.

\bigskip

Figure 8.3~ Comparison of the total cross-sections obtained in the two
extreme conditions of $\kappa=1$~ (pure confining case) and $\kappa=0$~
 (pure non-confining case).

\bigskip

\end{document}